# Surface Forces Apparatus measurements of interactions between rough and reactive calcite surfaces

Joanna Dziadkowiec*,[a], Shaghayegh Javadi[a,b,c], Jon E. Bratvold[d], Ola Nilsen[d], Anja Røyne[a]

[a] Physics of Geological Processes (PGP), The NJORD Centre, Department of Physics, University of Oslo, Norway

[b] The National IOR center of Norway, University of Stavanger, Norway.

[c] Petroleum engineering department, University of Stavanger, Norway.

[d] Centre for Materials Science and Nanotechnology (SMN), Department of Chemistry, University of Oslo, Norway

# ABSTRACT

Nm-range forces acting between calcite surfaces in water affect macroscopic properties of carbonate rocks and calcite-based granular materials, and are significantly influenced by calcite surface recrystallization. We suggest that the repulsive mechanical effects related to nm-scale surface recrystallization of calcite in water could be partially responsible for the observed decrease of cohesion in calcitic rocks saturated with water. Using the Surface Forces Apparatus (SFA), we simultaneously followed the calcite reactivity and measured the forces in water in two surface configurations: between two rough calcite surfaces (CC), or between rough calcite and a smooth mica surface (CM). We used nm-scale rough, polycrystalline calcite films prepared by Atomic Layer Deposition (ALD). We measured only repulsive forces in CC in $CaCO_3$-saturated water, which was related to hydration effects and/or roughness. Velocity-dependent adhesion forces were measured in monoethylene glycol (MEG) for relatively smooth surfaces in CC. Adhesive or repulsive forces were measured in CM in $CaCO_3$-saturated water depending on the calcite roughness, and the adhesion was likely enhanced by electrostatic effects. The pull-off adhesive force in CM became stronger with time and this increase was correlated with a decrease of roughness at contacts, which parameter could be estimated from the measured force-distance curves. That suggested a progressive increase of real contact areas between the surfaces, likely caused by gradual plastic deformation of calcite surface asperities during the repeated loading-unloading cycles. Reactivity of calcite was affected by mass transport across nm to μm-thick gaps between the surfaces. Major roughening was observed only for the smoothest calcite films, where the gaps between the two opposing surfaces were nm-thick over μm-sized areas, and led to a force of crystallization that could overcome confining pressures of the order of MPa. Any substantial roughening of calcite led to a significant increase of the repulsive mechanical force contribution.

# Introduction

Calcite is a crucial rock-forming, cement-forming and accessory mineral, significant in biomineralization and in the global $CO_2$ cycle. Nanometer-range, repulsive or attractive forces, acting between calcite surfaces in aqueous solutions, are critical to overall mechanical strength of calcite-bearing rocks, biomineralization processes, enhanced oil recovery (EOR) in chalk [1], and a range of industrial applications, in which calcite surface interactions play a major role. Measurements of such short-range forces could further explain possible mechanisms of $CaCO_3$ aggregation in natural biocomposite systems, such as nacre [2], and shed more light on properties of colloidal systems, in which granular calcite is used as an excipient, filler or a principal ingredient [3-6]. Engineering of $CaCO_3$ surface properties, and thus modification of surface forces, could produce functional carbonate fillers, significant for concrete, paper, plastics, and other materials constituting major amounts of carbonate additives [7]. Moreover, forces acting between the surfaces are intimately related to friction, where the contacts between the surfaces are usually limited to surface asperities, which, apart from mechanical deformation effects, experience attraction or repulsion during lateral motion of surfaces [8]. The presence of nanoscale grains, or asperities, on fault slip planes has been associated with low frictional resistance in carbonate rocks, which may potentially lead to fault weakening and earthquake triggering [9].

Apart from the fact that the forces between surfaces are severely affected by the properties of the surrounding liquid medium, such interactions, in the case of calcite, are further complicated by the reactivity and roughening of the mineral surfaces on exposure to water. This has major implications both for rocks, that undergo water-induced weakening, and for many colloidal systems. The water weakening phenomenon is related to a substantial loss of mechanical strength in fluid-saturated rocks. Early observations of this effect point to a mineralogy-dependence and a decisive contribution of processes occurring in fractures and at grain boundaries. Diffusion of reactive fluids into these solid-solid contacts amplifies subcritical crack growth, pressure solution and compaction, and thus contributes to a long-term creep deformation [10]. Observations of water-enhanced subcritical fracturing in poorly soluble materials have led to the conclusion that effects other than simple chemical dissolution must also act at solid interfaces. It was found that the decrease of solid surface energy due to water adsorption on mineral surfaces [11], or a preferential hydrolysis of strained mineral bonds [12] can significantly contribute to the weakening. Such effects may prevent adhesive interactions, and in turn lead to lower grain cohesion and fracture thresholds in rocks and materials [13-15]. The repulsive hydration effects related to water adsorption strongly depend on mineralogy, as well as on the chemical composition of the aqueous phase, since the surface hydration is greatly controlled by the solid-liquid interface structure [16-19].

Water weakening is found to be most severe in porous, sedimentary rocks, including carbonates [20]. Although substantial compaction in chalk reservoirs, induced by water injection [21], has been studied in-depth, the lingering question remains about the dominant mechanism causing the observed subsidence. Risnes, et al. [15] used mixtures of water and ethylene glycol to show that chalk strength decreases with increasing water activity of the pore fluid, and attributed the observed loss of cohesion mainly to repulsive forces due to water adsorption on calcite surfaces. Other mechanisms that may weaken water-flooded calcitic rocks include: destruction of intergranular capillary bridges [22], plastic strain increase [23], chemical dissolution at low stresses [24], pressure solution [15, 24-26], and enhanced grain-scale subcritical cracking [13]. Most studies conclude that coupled mechanisms must be in action, and link effects of water adsorption on calcite surfaces with the macroscopic properties of rocks [27-32].

Although the presence of repulsive forces between calcite surfaces in aqueous solutions has been recognized [15, 30, 33-34], the very nature of this interaction is only beginning to be understood. In recent experiments, nm-ranged forces between two calcite surfaces [35] or calcite-silica surfaces [36] have been directly measured by the Atomic Force Microscopy (AFM). In contrast to the adhesive interaction between freshly cleaved calcite surfaces in air and in ethylene glycol, Røyne, et al. [35] reported strong repulsive forces when the medium between the surfaces was $CaCO_3$-presaturated water. The magnitude of this force significantly exceeded the theoretical DLVO electric double layer repulsion [16], and was attributed to the hydration forces acting between the highly hydrophilic calcite surfaces. Such hydration repulsion was further resolved by Diao and Espinosa-Marzal [36] who evidenced the oscillatory nature of this force, a phenomenon related to the layering of water populated by counterions in different hydration states. Interestingly, the magnitude and onset of the repulsion were sensitive to the electrolyte concentration, which relates to the progressive dehydration of cations, squeezed in a confined film between the surfaces. These findings indicate that the nature of repulsive forces between calcite surfaces is indeed intimately related with the already well-established molecular details of a calcite-solution interface [37-40].

Surface roughness remains insufficiently addressed in the complex interactions between calcite surfaces. Water wettability of calcite surface has been found to increase with its roughness [41]. This observation was further emphasized by Chen, et al. [42] for EOR systems, in which dilute electrolyte solutions enhance oil desorption from calcite surfaces both by affecting colloidal forces and by increasing calcite surface roughness. It has been shown that even nanoscale details of surface topography significantly can affect the forces acting between the mineral surfaces [16, 43]. Roughness not only decreases the contact area between the surfaces, but also disturbs the arrangement of surface species, and generates a repulsive force due to elastic deformation of the highest asperities [44-45]. Moreover, calcite surfaces display strong, local variations of surface charge related to details of surface topography, as predicted by recent modelling studies [46]. Changes in roughness may thus severely affect forces between calcite surfaces, which are reactive and dynamic in contact with aqueous solutions [47]. Additionally, although varied in magnitude and range, both hydration forces and roughness often add an exponentially repulsive component to the total force acting between the surfaces, which complicates the interpretation of the measured forces [35, 44]. Here, we investigate the effect of surface roughness on forces between calcite surfaces by employing polycrystalline calcite substrates with multiple asperity contacts.

Even though the AFM is an extremely powerful tool in force measurements, it usually provides no information about the instantaneous changes in contact surface topography. In this work, we used the Surface Forces Apparatus (SFA), a force measuring technique that enables *in situ* observations of surface alteration processes by multiple beam interferometry (MBI) [42, 48-49]. We follow how the growth, dissolution and related changes in surface roughness in μm-sized contact areas affect the magnitude and range of forces between dynamic and rough, polycrystalline calcite surfaces. Moreover, we present a feasible setup for measuring interactions between calcite surfaces in the SFA, which technique is becoming more frequently used to study forces between various mineral surfaces [42, 50-54].

# Materials and Methods

## Preparation of Calcite Films

Thin, polycrystalline films of calcite were grown by the atomic layer deposition (ALD) method using a commercial F-120 Sat reactor from ASM Microchemistry. The process was adapted from Nilsen, et al. [55] with Ca(thd)$_2$ (Volatec; 97 %; Hthds = 2,2,6,6-tetramethylheptan-3,5-dione) and ozone as reaction precursors. To ensure deposition of phase-pure and crystalline CaCO$_3$, CO$_2$ (Praxair; 99.7 % pure) was pulsed after O$_3$, in line with the findings of Nilsen, et al. [55]. A constant carrier gas flow was provided from bottled N$_2$ (Praxair; 99.999 %). Ca(thd)$_2$ was kept at 195 °C to ensure sufficient sublimation, and ozone was generated by feeding O$_2$ (Praxair; 99.5 %) into an ozone generator (In USA AC series) producing ca. 15 % (200 g/N·m$^3$) O$_3$ at a flow of ca. 500 sccm. The deposition temperatures ranged from 250 to 350 °C. Thicknesses and refractive index values (at $\lambda$ = 632.8 nm) were investigated using a spectroscopic ellipsometer (J. A. Woollam alpha-SE), fitting the data to a Cauchy model (CompleteEASE software) for transparent films. Additional modelling was performed using the SFA coupled with MBI in the open source Reflcalc software [56] as described below. The thickness of the films varied between 100 and 200 nm, depending on the number of cycles used for the deposition. The detailed deposition parameters are given in the Supplementary Materials (SM).

## Preparation of SFA Samples

To facilitate the SFA measurements, which require a µm-range sample thickness and a semi-reflective metal layer present on the back side of the sample, the ALD calcite films were deposited on mica substrates **(Figure 1A)**. The mica substrates were first freshly cleaved into uniformly thick (1–10 µm) layers [57], placed on a freshly-cleaved mica backing sheet, and coated with a 45 nm-thick layer of Au (with the E-beam Leybold L560 evaporator). Subsequently, these Au-coated pieces of mica were flipped over to expose the Au-free surface, and were placed on a freshly cleaved mica backing sheet (15 x 5 cm$^2$) with the Au surface facing down. To avoid any glue or tape in the ALD reaction chamber, the pieces of Au-mica were clamped down on each end with stripes of freshly cleaved mica, which adhered to the mica backing plate. The use of standard Ag-coated mica or mica that was pre-attached on the SFA disks with epoxy glue was not feasible due to oxidation of Ag or epoxy decomposition inside the ALD reaction chamber at temperatures reaching up to 350 °C. The Au-free mica surface was exposed to air just before the deposition of the calcite films, and the sample was handled in a laminar flow cabinet at all times. After calcite deposition the obtained calcite/mica/Au films were glued with EPON 1004F thermal glue to standard SFA cylindrical disks (SurForce LLC; R = 0.02 m). Alternatively, the calcite films were deposited on the standard SFA glass disks **(Figure 1B)** or flat glass slides for characterization, both previously coated with 4 nm thick Ti and 45 nm thick Au layers. Ti was used to improve the adhesion of Au to the glass surface. The initial rms roughness of the Ti/Au layer on glass was 0.8 nm (scan size 5x5 $\mu m^2$), as characterized by AFM. Additionally, the setup with calcite deposited directly on the Au-coated SFA disks, did not allow measurements in the symmetric system (calcite-calcite), due to insufficient thickness of calcite layer, and was only used for measuring mica-calcite interactions. For these measurements, bare mica surfaces were prepared as described by [57]. Pieces of freshly cleaved, optical-grade mica sheets (S&J Trading Inc., USA), with a thickness of 1–10 μm, were cut using a hot platinum wire under a laminar flow, and back-silvered (55 nm) with a thermal evaporator (Balzers BAE 250).

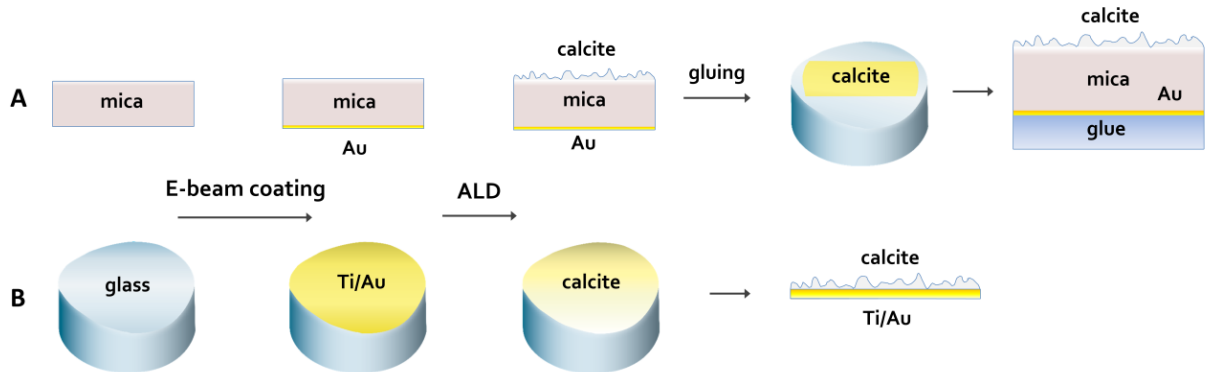

**Figure 1.** *Scheme showing two routes of calcite films preparation for the SFA measurements on the standard cylindrical SFA disks with the radius of curvature R = 0.02 m. E-beam coating was used to deposit metal layers (Au – gold and Ti – titanium); ALD was used to deposit calcite layers. A. calcite deposited on mica substrate; B. calcite deposited on Au-coated glass disks.*

## SFA Measurements

The Surface Forces Apparatus (SFA) is a force measuring technique that enables a direct measurement of the forces acting between two macroscopic surfaces as a function of the distance between them, and has been described in detail elsewhere [58-62]. Briefly, the SFA is based on a simple concept of approaching and separating two cylindrical surfaces, one of which is attached to a force-measuring spring. The distance between the surfaces is controlled independently in two ways: by applying a known displacement mechanically ($D_{applied}$), and with an optical interferometric technique ($D_{measured}$). When there is a force F(D) acting between the surfaces, it will deflect the force-measuring spring. The separation values measured by interferometry are then not equal to the applied separation values, and the force can be calculated as:

$$F(D) = k(D_{applied} - D_{measured}) \qquad \text{(Eq. 1)}$$

where *k* (N/m) is the spring constant, calibrated by applying known weights to the SFA spring and measuring the displacement (here $k = 2 \cdot 10^3$ N/m). In order to enable comparison with force measurements in different systems through the Derjaguin approximation [63], the force is usually normalized with the local radius of curvature *R* of the surfaces, which is measured from the shape of FECO fringes (SM; **Figure S9**). The great benefit of the SFA is a possibility to observe changes in surface topography *in situ*, which are indicated by a shape of the FECO fringes [48], here with a lateral resolution of 0.624 µm/pixel. In addition, the SFA uses relatively large contact areas (50–150 µm in diameter), which enables us to study influence of confinement on surface recrystallization, where nm-thick liquid films are trapped between the contacting surfaces over µm-sized areas.

The force measurements were performed using the Surface Forces Apparatus (SFA2000) supplied by SurForce LLC, USA [62], equipped with a spectrometer (Princeton Instruments IsoPlane SCT320 with a PIXIS2048B camera) and a camera (Thorlabs DCC1645C) for surface topography observations (resolution of 0.015 µm/pixel). The spectrometer was calibrated in the wavelength (λ) range 520–630 nm, using a Hg light source (Princeton Instruments, Intellical). Spectrometer gratings with different resolutions and a λ range (600, 1200 or 1800 g/mm), and at varied center λ, were used depending on the sample thickness. Reference images of the surfaces in contact were captured using the 'step and glue' mode, in the full λ range 520–630 nm (Supplementary Information; S2.1). The SFA data was analyzed by simulating the FECO fringe patterns using the open source Reflcalc software [56], which uses the matrix method for stratified samples adapted from Schubert [64]. The wavelength positions of the FECO fringes were identified using the MATLAB software. All details of the SFA data analysis are described in SM.

## Characterization of Calcite Films

Atomic Force Microscopy (AFM) was used to measure the topography of ALD calcite films. We used a JPK NanoWizard®4 Bioscience AFM, in QI-mode, with a ContAl-G cantilever (NanoSensors, $k$ = 0.2 N/m and $l$ = 450 µm). The initial roughness of the surfaces was analyzed in air, within maximum 2 weeks after each deposition (all the samples were stored in a desiccator cabinet). Several samples were used to follow the evolution of roughness for up to 3 hours in $CaCO_3$-saturated solutions in a ~3.5 ml fluid cell, semi-sealed with a silicon ring (under stationary fluid conditions, with no fluid exchange during each measurement). The final topography of the samples used for the SFA experiments was measured after drying them gently with $N_2$ or in air. The AFM was kept in a temperature-controlled enclosure, at 24.5 ±0.5 °C. The scan size varied between 0.5x0.5 and 60x60 µm² to account for variation in size and density of asperities, and the samples were scanned in several positions. Surface roughness data was characterized by root mean square (rms) values of the height data, calculated as $rms = \sqrt{\frac{1}{N}\sum_{i=1}^{N} y_i^2}$, where $N$ is the number of height data points and $y$ is the measured height in each point.

To reveal which $CaCO_3$ phase was deposited on mica, two film samples of ~200 nm thickness, deposited at 250 or 300 °C, were investigated with the Raman spectroscopy, using a Jobyn–Yvon Horiba T64000 instrument. The spectra were collected in the wavenumber range between 843 and 1338 cm$^{-1}$, using the laser operating at 787 nm, and with the spectral resolution of 2.0 cm$^{-1}$ (grating 900 lines/mm). For each measurement, five scans of 120 s were averaged, and background fluorescence was corrected by fitting polynomial functions. The spectra were calibrated using two peaks of the paracetamol standard at 857.9 cm$^{-1}$ and 1323.9 cm$^{-1}$.

X-ray diffraction (XRD) analysis was performed on calcite films deposited on Au-coated glass slides, with a Bruker AXS D8 Discover powder diffractometer in Bragg-Brentano configuration, equipped with a Lynxeye detector, using Cu Kα1 radiation and a Ge(111) monochromator.

Scanning Electron Microscopy (SEM) was performed using Hitachi SU5000 FE-SEM to observe the initial and final morphology of samples, which were gently dried with $N_2$, immediately after the SFA experiments. The samples were coated with Au, and images were collected using secondary electrons (SE) at accelerating voltage of 10 kV or 15 kV.

## Preparation of Solutions

$CaCO_3$-saturated solutions used in the SFA and AFM were prepared by presaturating deionized water (Milli-Q, 18.2 MΩ) for a week or more with an excess amount (~1 g/L) of synthetic, ultrapure calcite powder (Merck KGaA), which was baked at 300 °C for 2 hours, prior to use. All solutions were filtered with 0.45 µm polyether-sulfone Macherey-Nagel syringe filters, prior to use in the SFA. pH of the solutions was measured with the S470 SevenExcellence™ Toledo Mettler instrument, calibrated with pH 4, 7 and 10 buffers. The concentration of $Ca^{2+}$ in the solutions was measured by AAS (Perkin Elmer AAnalyst 400 Atomic Absorption Spectrometer), with 0.4 ml of 10% $LaCl_3$ per 4 ml of a water sample ($LaCl_3 \cdot H_2O$, Prolabo Rectapure®, 99.99%) used as a $Ca^{2+}$ release agent. Calcite saturation index (SI) values were calculated with PHREEQC software (phreeqc database) using the AAS measured $Ca^{2+}$ concentration and pH values, assuming atmospheric $CO_2$ partial pressure of $pCO_2 = 3.5$ ($10^{-3.5}$ atm), in the case of open systems and $pCO_2 = 10^{-6.2}$ atm for the closed systems. Ethylene glycol (monoethylene glycol (MEG); ethane-1,2-diol; Merck, reagent grade, 99.5% pure) was used as supplied. Each time a solution was injected into the SFA chamber, the chamber was rinsed with an excess amount (~150 ml) of the injected solution to reduce any possible contamination. Solutions were usually injected in the SFA with the calcite surfaces brought into close contact, to limit their fast dissolution upon equilibration. Experiments with ethylene glycol were performed using only a liquid droplet injected in between the separated surfaces, using a needle and a syringe. The droplet was exchanged multiple times to ensure a complete surface wetting.

## DLVO and Roughness Modelling

Van der Waals (VdW) and electric double layer forces (EDL) were calculated according to the DLVO theory [65-66], both in symmetrical systems: calcite-calcite and mica-mica, and in the asymmetric calcite-mica system, using the equations for two cylindrical surfaces of radius R, adapted from Israelachvili [16]. In all cases, VdW forces were calculated as:

$$F_{VdW} = \frac{-A \cdot R}{6 \cdot D^2} \ (N) \qquad (Eq.2)$$

where A is the Hamaker constant, and D is the distance between the surfaces. The values of non-retarded Hamaker constants calculated with the full Lifshitz theory, were adapted from literature data [67]: $1.34 \cdot 10^{-20}$ J for two mica surfaces across water, $1.44 \cdot 10^{-20}$ J for two calcite surfaces across water, and $1.35 \cdot 10^{-20}$ J for asymmetric mica and calcite surfaces across water. Although the experimental Hamaker constant value of $2.2 \cdot 10^{-20}$ J for two mica surfaces across water is more frequently used [59], for the sake of comparison between the three systems we used the Hamaker constant values all computed with the same mathematical representation (SNP method), which additionally uses improved spectral parameters of water [67].

EDL forces in the symmetrical systems, in $CaCO_3$-saturated solutions, were estimated according to the approximate expression for mixed electrolytes at a constant surface potential $\psi_0$ of isolated surfaces:

$$F_{edl}(D) = 2\pi R \varepsilon_0 \varepsilon \kappa \psi_0^2 e^{-\kappa D} \ (N) \qquad (Eq.3)$$

where $\kappa^{-1}$ is the Debye length calculated as $\kappa = \sqrt{\sum_i \frac{C_i e^2 z_i^2}{\varepsilon_o \varepsilon kT}}$ for each ion species *i* in the electrolyte, *R* is the radius of the curvature, *D* is the distance between the surfaces, $\varepsilon_0$ is electrical permittivity of vacuum, $\varepsilon$ is the water dielectric constant, *k* is the Boltzmann constant, *T* is the absolute temperature, $C_i$ is the bulk concentration and $z_i$ is valency of ion species i in the electrolyte. To estimate Debye length values, the concentration of ions in the $CaCO_3$-saturated solutions, which are mixed 2:1 electrolytes at pH values ~ 9 ($HCO_3^- \gg CO_3^{2-}$), was calculated in the PHREEQC software, using the AAS-measured concentration of $Ca^{2+}$ and initial pH values of the solutions, assuming a closed system ($pCO_2 = 10^{-6.2}$ atm).

EDL forces in the asymmetric system, between calcite and mica surfaces, were estimated using the Poisson-Boltzmann equation for dissimilar surfaces of low surface charge, adapted from Trefalt, et al. [68] and Diao and Espinosa-Marzal [36]:

$$F_{edl}(D) = 2\pi R \varepsilon \varepsilon_0 \kappa \frac{2\psi_m \psi_c e^{-\kappa D} + e^{-2\kappa D}((2p_m-1)\psi_m^2 + (2p_c-1)\psi_c^2)}{1-(2p_m-1)(2p_c-1)e^{-2\kappa D}} \quad \text{(Eq.4)}$$

where $\psi_m$ and $p_m$ are mica surface potential and regulation parameter, $\psi_c$ and $p_c$ are calcite surface potential and regulation parameter; $p = 1$ for a constant surface charge of the surfaces and $p = 0$ corresponds to a constant surface potential. The mica surface potential, $\psi_m = -70$ mV, was estimated for two mica surfaces in $CaCO_3$-saturated solutions (SM, **Figure S13**). The calcite surface potential, $\psi_c$, was assumed to be -20 mV [69], and the calcite regulation parameter $p_c = 0.62$ was adapted from [36]. The mica regulation parameter $p_c = 0.1$ was determined by fitting EDL forces between two mica surfaces in $CaCO_3$-saturated solutions at a previously determined surface potential $\psi_m = -70$ mV, using Eq. 3. The values determined this way are only roughly estimated, however, should at least be in the correct range.

The contribution of roughness to the total force between the surfaces was estimated using a model adapted from Parsons, et al. [45], which considers both a roughness-averaged non-contact force between the surfaces and a repulsive contact force due to elastic deformation of the highest surface asperities. The first component of the model uses a probability distribution of surface heights to account for roughness, which perturbs the non-contact force acting between the surfaces. The distribution of surface heights was represented with histograms, based on the AFM roughness measurements of the deposited calcite films, and the roughened force between the surfaces was calculated using the formula:

$$F_{non-contact}(D) = 2\pi R * \frac{1}{N_1 N_2} * \sum_i \sum_j H_{1i} * H_{2j} * G(h_{2j} - h_{1i} - \bar{h}_2 + \bar{h}_1 + D) \quad \text{(Eq.5)}$$

where $N_1$ and $N_2$ are normalization factors ($N_1 = \sum_i H_{1i}$; $N_2 = \sum_j H_{2j}$), $H_{1i}$ is a histogram of heights $h_{1i}$ of the first surface, with a mean height value $\bar{h}_1$, and $H_{2j}$ is, accordingly, the histogram of the second surface, and $G$ is the DLVO interaction energy for smooth surfaces, at a roughness-averaged separation. The second component introduces a repulsive contact force due to elastic compression of asperities:

$$F_{contact}(D) = \frac{4RE_r\sigma_m}{15\sqrt{\pi}} \sqrt{\frac{\sigma_m}{r_a}} \exp\left(-\frac{D^2}{4\sigma_m^2}\right) * f\left(\frac{D}{\sigma_m}\right) \quad \text{(Eq. 6)}$$

where $E_r$ is a reduced Young's modulus ($\frac{1}{E_r} = \frac{1-v_1^2}{E_1} + \frac{1-v_2^2}{E_2}$) with Young's modulus $E$ and Poisson's ratio $v$ of surface 1 and 2, respectively; $r_a$ is an average reduced radius of surface asperities ($\frac{1}{r_a} = \frac{1}{r_1} + \frac{1}{r_2}$) of surface 1 ($r_1$) and 2 ($r_2$); $\sigma_m$ is the mean rms roughness of the surfaces ($\sigma_m = \sqrt{\sigma_1^2 + \sigma_1^2}$); and $f(x) = \sqrt{x}[(1 + x^2)K_{0.25}\left(\frac{x^2}{4}\right) - x^2 K_{0.75}\left(\frac{x^2}{4}\right)]$, with $K_n$ representing two modified Bessel functions of the second kind with $n = \frac{1}{4}$ and $n = \frac{3}{4}$.

The final 'roughened' force between the surfaces was calculated as the sum of two contributions:

$$F_{rough}(D) = F_{non-contact}(D) + F_{contact}(D) \quad \text{(Eq.7)}$$

The used values of mica surface rms roughness (0.05 nm), Young's modulus (E = 34.5 GPa) and Poisson's ratio (v = 0.205 (-)), were adapted from [70]. The roughness of the calcite films was determined by AFM, using the film samples before and after the SFA experiments, and the values of Young's modulus and Poisson's ratio were 72.35 GPa and 0.3 (-), respectively. The average radius of asperities was estimated above an arbitrarily chosen height threshold values from the AFM topography scans.

## Results and Discussion

### Characterization of Calcite Films

SEM images indicated that the polycrystalline calcite films deposited at 300 °C on mica were initially composed of triangular, nm-sized (<50 nm) crystals, the morphology of which could not be precisely resolved with SEM. The films were continuous over cm-sized areas, with the presence of bigger, µm-sized, polycrystalline aggregates on some of the surfaces **(Figures 2A, B)**. These aggregates were composed of overgrown crystals with a rhombohedral morphology characteristic for calcite, with dominant {104} bounding faces. The presence of aggregates was common, but in varied amounts between each deposition, giving rise to a poorly controlled roughness of the films.

Due to the overlap of muscovite mica diffraction peaks with the weak intensity $CaCO_3$ peaks, XRD patterns were measured for $CaCO_3$ films grown on Au-coated glass slides. In most cases, the deposited $CaCO_3$ phase was calcite. The crystallinity and orientation, and most likely the morphology, of the calcite films on the Au substrate varied with the deposition temperature **(Figure 3)**. At 250 °C, XRD revealed two distinct grain orientations with (006) or (104) calcite planes parallel to the substrate. At 300 °C, most of the deposited films were (104)-oriented, with some samples having additionally (006)-oriented grains. At 350 °C, (104) orientation was prevailing. Additional, low-intensity peaks were measured at 2θ of 20.9° and 42.7° for a few samples, which could be most likely be attributed to (00$l$) reflections of vaterite (SM, **Table S3; Figure S10**). Moreover, it is possible that a certain amount of amorphous $CaCO_3$ was present in the crystalline films, the quantity of which was likely to decrease with the deposition temperature, as suggested by Nilsen, et al. [55]. The broad background features between 20° and 30° 2θ were due to the $SiO_2$ glass substrate. Only the films with distinct calcite XRD diffraction peaks were used in the SFA experiments, however the peaks intensity could be affected by a presence of µm-sized aggregates in the films.

As the substrate can influence the orientation, crystallinity [55], and perhaps the phase of the $CaCO_3$ films, it was possible that the films deposited on mica would yield a different $CaCO_3$ phase than those deposited on gold. Raman spectroscopy measurements of the mica-deposited films revealed a presence of a single band with a shift centered at 1088 or 1089 cm$^{-1}$ (for the films deposited at 250 and 300 °C, respectively; **Figure 3**). This shift corresponds to the most intense vibration in the symmetric stretching mode of the carbonate groups v1 $(CO_3)^{2-}$, and could be attributed to calcite or, less likely, aragonite. The typical v1 shift for a pure calcite phase varies between 1088 and 1085 cm$^{-1}$, whereas this vibration occurs at slightly lower wavenumbers for aragonite, at 1084–1086 cm$^{-1}$ [71-75]. Nevertheless, the rhombohedral morphology observed with SEM **(Figure 2A)** was typical for calcite and thus the presence of aragonite was unlikely. In addition, Raman observations did not evidence any presence of vaterite, which can be identified by a characteristic splitting of the v1 band at 1090 and 1075 cm$^{-1}$ [76], nor amorphous $CaCO_3$, which presence would be indicated by a broad Raman carbonate v1 band centered at approximately 1077–1082 cm$^{-1}$ [77]. The latter, however, could be related to insufficient thickness of the calcite films (200 nm).

The thickness, thickness gradients, detailed deposition parameters and XRD patterns of all deposited films are provided in SM (**Tables S2, S3; Figure S10**).

The topography of the calcite films, deposited both on mica and Au-coated glass substrates, at 250, 300 and 350 °C was measured with the AFM in air. The detailed parameters are given in SM (**Tables S8, S9**). **Figure 4** shows the rms roughness values as a function of scan size ($ss$), for the films before and after the SFA experiments. In general, we observed that the rms roughness varied within 2 orders of magnitude, and it increased with the increasing scan size because of the presence of larger asperities (<1 µm) on the surface. The films deposited on mica exhibited smaller rms values (0.9 nm ($ss$ = 0.2 µm) $\leq$ rms $\leq$ 109 nm ($ss$ = 40 µm)) than the ones on the Au-coated substrates (3 nm ($ss$ = 0.1 µm) $\leq$ rms $\leq$ 202 nm ($ss$ = 5 µm)), which is related to the higher initial roughness of the Au-substrates. On average, the films grown at 250 and 350 °C were smoother (max rms = 8.0 nm, $ss$ = 10 µm; rms = 17 nm, $ss$ = 20 µm, respectively) than the ones at 300 °C (max rms = 202 nm, *ss* = 5 µm). However, we prepared most of the films at 300 °C and we measured small rms values for this temperature as well. In general, the distribution of surface asperities was not homogenous, and the maximum radius of the highest asperities was estimated to be 0.5 µm for the roughest surfaces. The smoothest surfaces (rms <3 nm) had asperities with an average radius of ~10 nm.

Interestingly, the rms of the films after the SFA experiments (**Figure 4,** filled symbols) in $CaCO_3$-solutions did not show a large variation with the scan size as the films before the SFA, and they were in the higher range of all the measured values. The films after the SFA experiments in MEG were generally smoother than the ones in $CaCO_3$-solutions (max rms = 59 nm, *ss* = 50 µm; rms = 229 nm, *ss* = 30 µm, respectively).

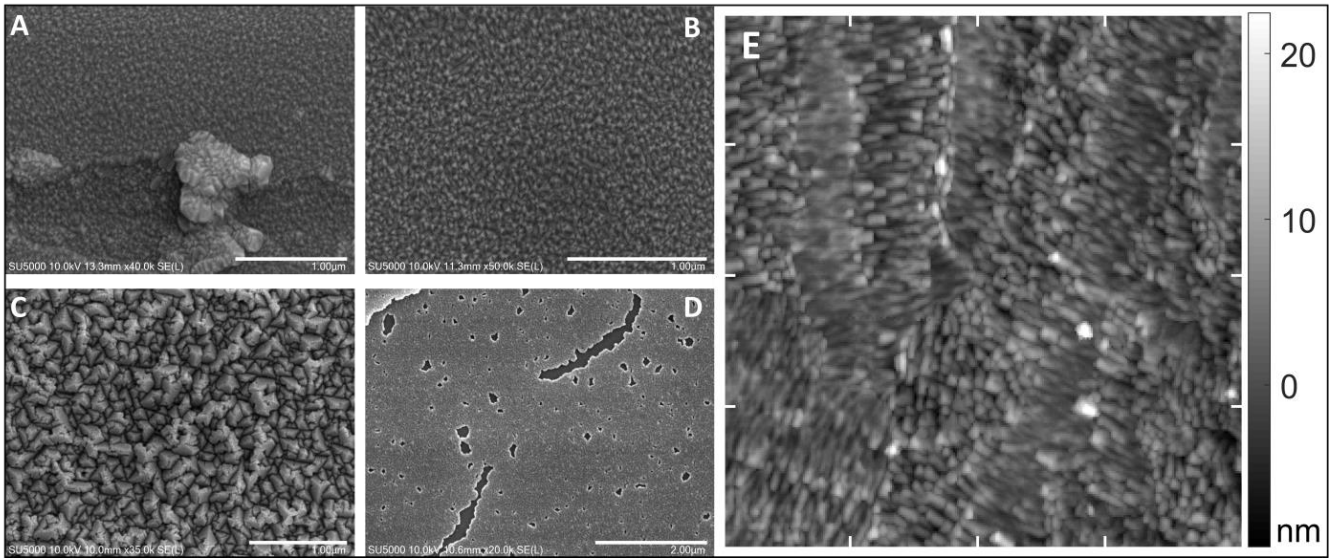

**Figure 2.** *SEM SE images of the initial (A, B) and final (C, D) topography and AFM height measurement (E) of polycrystalline calcite films deposited on mica (A, B, C, D) and gold (E) substrates: A. 1 µm size asperities, present on some of the surfaces, were polycrystalline (BAJ1041, 300 °C); B. average size of crystals was smaller than 50 nm (BAJ1041, 300 °C); C. region with roughened crystals after the SFA experiment (CM170711; BAJ2059); D. region with dissolved areas after the SFA experiment (CC170713; BAJ2059). E. AFM of the 200 nm-thick film deposited at 250 °C (scan size 4x4 µm, ticks every 1 µm, rms = 3.9 nm, BAJ1068). The scale bar is 1 µm (A, B, C) or 2 µm (D).*

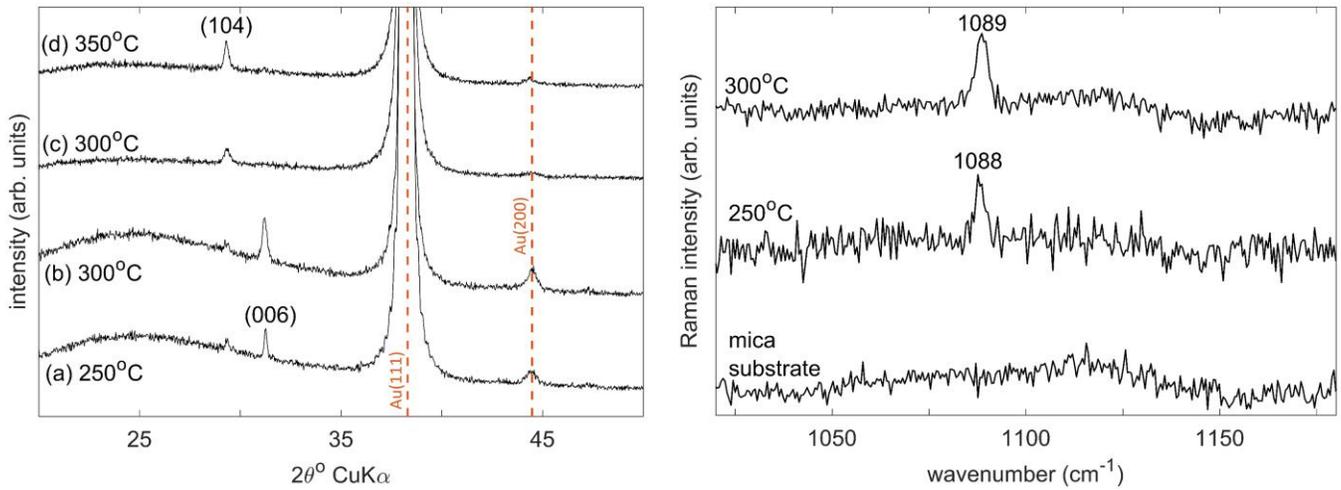

**Figure 3.** *Left: XRD of the chosen calcite films deposited at 250, 300, and 350 °C on Au-substrate: samples a. BAJ1068 (200 nm), b. BAJ1005 (200 nm), c. BAJ1041 (100 nm) and d. BAJ2059 (100 nm). Right: Raman spectra of a bare mica substrate and calcite films deposited on mica at 250 and 300 °C (samples BAJ1068 and BAJ1005, respectively).*

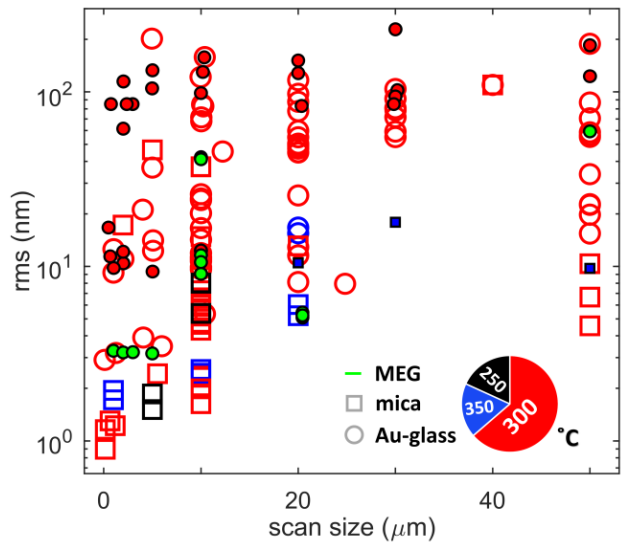

**Figure 4**. *AFM-measured roughness of calcite films as a function of a scan size. Blue, red and black colors correspond to calcite film deposition temperature. The empty, larger sized symbols represent measurements in air after the film deposition (max. 2 weeks). The smaller, filled symbols represent measurements for the calcite surfaces used in the SFA experiments (in $CaCO_3$-saturated solutions and in MEG – indicated in green), after drying the surfaces with $N_2$ or in air. □ – films deposited on mica substrates; or ○ – on Au-coated glass substrates.*

## Effect of Initial Crystal Morphology on the Reactivity of Calcite Films

Reactivity of the polycrystalline calcite films, deposited at three different temperatures (250, 300, and 350 °C), was followed in the AFM (SM; **Figure S22**) and in the SFA, upon injection of $CaCO_3$-saturated solution (**Figure 5;** solution parameters in SM, **Table S7**).

In the AFM, where the measurements were performed with single, unconfined calcite surfaces, we did not evidence any significant recrystallization or formation of larger sized crystals for any of the samples within 1 h. The roughness of the films increased slightly with time and the highest increase in rms was 24 nm (BAJ1005, 300 °C). The initial and final topography of the samples are shown in SM; **Figure S22**. We additionally followed changes in roughness for one sample (300 °C, BAJ2046) in air over 2 months and we did not evidence any significant recrystallization (data not shown).

In the SFA, the equilibration process was followed for dissimilar calcite and mica surfaces in confinement and involved the following steps: the surfaces were initially contacted in air; several loading-unloading cycles (FR) in air were performed (~1 h); the surfaces were brought into a close contact again; and the $CaCO_3$-saturated solution was injected. During the injection, the surfaces were kept in contact under a constant, moderate applied load (<850 mN/m, corresponding to ~5.5 MPa for two smooth surfaces with a contact area radius of 100 µm). We observed that the films grown at 250 °C underwent significant recrystallization, whereas the depositions at 300 and 350 °C yielded much more stable films.

The much more pronounced reactivity of the calcite films deposited at 250 °C was manifested in the SFA in three ways: (1) we observed that ~1 µm-sized crystals were present in the calcite layer when the surfaces were contacted in air for the first time **(Figure 5C)**; (2) the thickness of calcite was progressively decreasing during FR in air, suggesting a presence of a water film between the surfaces; (3) this relatively fast dissolution continued when the $CaCO_3$-saturated solution was injected, until much larger (>5 µm) calcite crystals rapidly formed everywhere on the calcite sample and replaced the polycrystalline film **(Figure 5D)**. Such major recrystallization suggested an Ostwald ripening process, driven by the surface free energy minimization [78]. A concurrent AFM analysis of a single, unconfined 250 °C calcite surface, showed no evidence of the ~1 μm-sized aggregates prior to the injection, and no growth of the bigger ~5 μm-sized crystals in the solution. SEM observations revealed that this AFM calcite film remained intact over large areas (in contrast to the SFA sample) and only scarce bigger calcite crystals were present on the sample after the experiment (~50 crystals/mm$^2$ in comparison to ~3000 crystals/mm$^2$ for the SFA sample; CC160429; SM, **Figures S15, S16**).

The lack of the ~1 μm-sized aggregates in air as observed with the AFM, suggested that the calcite films deposited at 250 °C could initially recrystallize upon placing them in a close contact (load ~100 mN/m) with mica in air, perhaps due to a capillary condensation effect that can occur between samples in a crossed cylinders SFA geometry (R = 2 cm) [79-80]. Additionally, these ~1 μm-sized crystals were mostly scattered around the contact region **(Figure 5C)**. A similar distribution of crystalline deposits, nucleated via capillary condensation, has been previously evidenced in the SFA, with the crystals growing around the contact but not in the very center of it [81-82]. However, the FECO patterns did not indicate a presence of a continuous capillary bridge between the surfaces in air (which should be visible due to differences in refractive indices of water and air). Such continuous capillary bridge would be also unlikely to persist through the duration of the experiment because the surface separation exceeded 1 μm (nm-range separations were only established in places were big 1 μm-sized asperities were touching the mica surface). It is known that cleaved {104} calcite surfaces readily adsorb Å-thick water films in air [83-84], irrespective of the humidity level. Even such thin water films can induce recrystallization of calcite, however, this also occurs on an Å-scale [37, 85], and the recrystallization that we observe results in features that are orders of magnitude larger. The adsorbed water film could be much thicker on rough surfaces [40], because the wettability of hydrophilic surfaces increases with their roughness [41, 86]. However, if the water film on isolated surfaces was sufficient for the development of 1 μm-sized crystals due to recrystallization, we would expect to see the same phenomenon when imaging the sample surfaces in air with the AFM.

This was not the case. We therefore suggest that for the rough 250 °C calcite films, multiple, small capillary bridges formed on the highest asperities in contact with mica, and that this led to growth of 1μm-sized crystals in air in these contact places. Either the bridges were too small to resolve with the SFA (resolution = 0.624 μm/pixel), or the growth happened fast, before we started to image the sample, as no further recrystallization events occurred during the FR measurements in air when the samples were repeatedly separated and brought to contact.

Although no more μm-sized crystals precipitated in the contact region during the subsequent FR in air, the calcite surfaces deposited at 250 °C were progressively dissolving. This further indicated that the adsorbed water layer was present on the surfaces. The measured decrease in calcite thickness in air was approximately 0.2 nm/s (CM160511), and was rather constant throughout the FRs (for ~1.2 h), irrespective of whether the surfaces were loaded or unloaded, excluding a decisive contribution of mechanical deformation effects. This dissolution rate significantly dropped after purging the SFA chamber with $N_2$, causing a partial drying of the surfaces. These results suggest that the adsorbed water film was thick enough to induce the dissolution of the calcite layer in air, but the surfaces had to be kept in contact for longer times for the major reprecipitation to occur, as at the beginning of the experiment.

The dissolution of the films deposited at 250 °C continued in a $CaCO_3$-saturated solution until the calcite films completely dissolved and large (>5 μm), overgrown calcite crystals, with dominant {104} faces, precipitated on the surface (**Figure 5D;** SM, **Figure S15**). The appearance of the bigger crystals was indicated by discontinuities in FECO fringe patterns, which developed due to the new crystal faces scattering the white light directed through the surfaces **(Figure 5B)**. Observation of the changing FECO patterns and the irregular shape of the newly precipitated crystals, reflected their fast growth, that occurred after the injected solution wetted the calcite films (the biggest crystals appeared in the contact region 20 minutes after the injection for CM160429; 2 and 40 minutes for CM160511). At first, it seemed surprising that the >5 μm crystals, that formed after injection of the solution, appeared everywhere on the SFA disk, and not in the AFM experiments. If the confinement, and resulting supersaturation in the gap, induced their growth, we would expect them to form only around the SFA contact region where the separation between the surfaces was the smallest. Recent experiments have shown that the precipitation of $CaCO_3$ could have been altered in a confined pore as big as 10 μm [87], but we found these particles even at the edges of the SFA sample, where the distance to the mica surface had been on the order of 1 mm. Most likely, momentary concentration gradients close to the calcite surface were present (even in such a big gap) in a stationary solution in the SFA as the calcite film was progressively dissolving, which is supported by the fact that the big crystals did not nucleate immediately after the injection. In contrast, the movement of an AFM cantilever could equal the concentrations on scanning, making the saturated layer close to the surface much thinner. In the previous AFM measurements of single calcite surfaces in stagnant solutions, high flux of ions leaving the dissolving samples has been reported to cause higher saturation in a boundary layer near the sample surface, and as such, even with a single, unconfined surface, dissolution and growth could have been transport-limited [88-90]. We thus suggest that a much thicker boundary layer could develop in the SFA, mainly because the dissolution rate of the calcite films deposited at 250 °C was much faster in comparison with the films grown at higher temperatures, but also due to the limited diffusion out of the gap in the normal direction, which was constrained by the two surfaces. Therefore, the recrystallization of the most reactive calcite films grown at 250 °C seemed to be primarily driven by the minimization of the surface free energy, but the confined SFA geometry also contributed to this process by enhancing the supersaturated conditions that developed in the boundary layer close to the calcite surface.

Additionally, our results do not suggest that the surface chemistry of mica primarily influenced the dissolution and major reprecipitation of these most reactive calcite films deposited at 250 °C, since the films dissolved and recrystallized in the same fashion in a symmetric configuration with 2 opposing calcite surfaces (see also *Effect of Dissimilar Surfaces*).

The calcite films grown at 300 °C were more stable both in air and upon exposure to the $CaCO_3$-saturated solution, and did not undergo any abrupt recrystallization. Similarly, no significant changes nor ripening occurred for the films deposited at 350 °C, despite the initial presence of some bigger aggregated calcite crystals (<1 μm) on the surface **(Figures 5K, L; Figure S22D).** However, it is possible that the films deposited at 300 °C were partially composed of the more reactive crystals typical for the 250 °C films, as suggested by similar XRD patterns for these two deposition temperatures in some cases (**Figure 3**).

As such, the reactivity of the ALD calcite films in $CaCO_3$-saturated water depended on the film deposition temperature. This was most likely related to differences in the crystal morphology, as suggested by different orientation of the calcite crystals grown at the lower temperature (prevailing (006) planes parallel to the substrate; **Figure 3**; SM, **Figure S10**). The films deposited at 300 and 350 °C, with dominant (104) planes parallel to the substrate, did not undergo any fast dissolution and subsequent recrystallization, as they were most likely composed of crystals bounded by more stable, low-energy {104} planes. The ALD deposition temperature influences the growth dynamics of calcite films and thus controls which sets of planes terminate the calcite crystals. Nilsen, et al. [91] has previously observed that calcite films deposited on different substrates at 250 °C have been composed of crystals most likely bounded by a {108} planes family, and, dominantly, a {104} type at higher temperatures, which suggests a potentially higher reactivity of the crystals deposited at 250 °C and their recrystallization driven by lowering the surface free energy (Ostwald ripening). In our data, this was further supported by visible differences in morphology of the samples grown at 250 and 300 °C as indicated by AFM and SEM data **(Figure 2)**. However, the set of bounding planes could not be precisely identified.

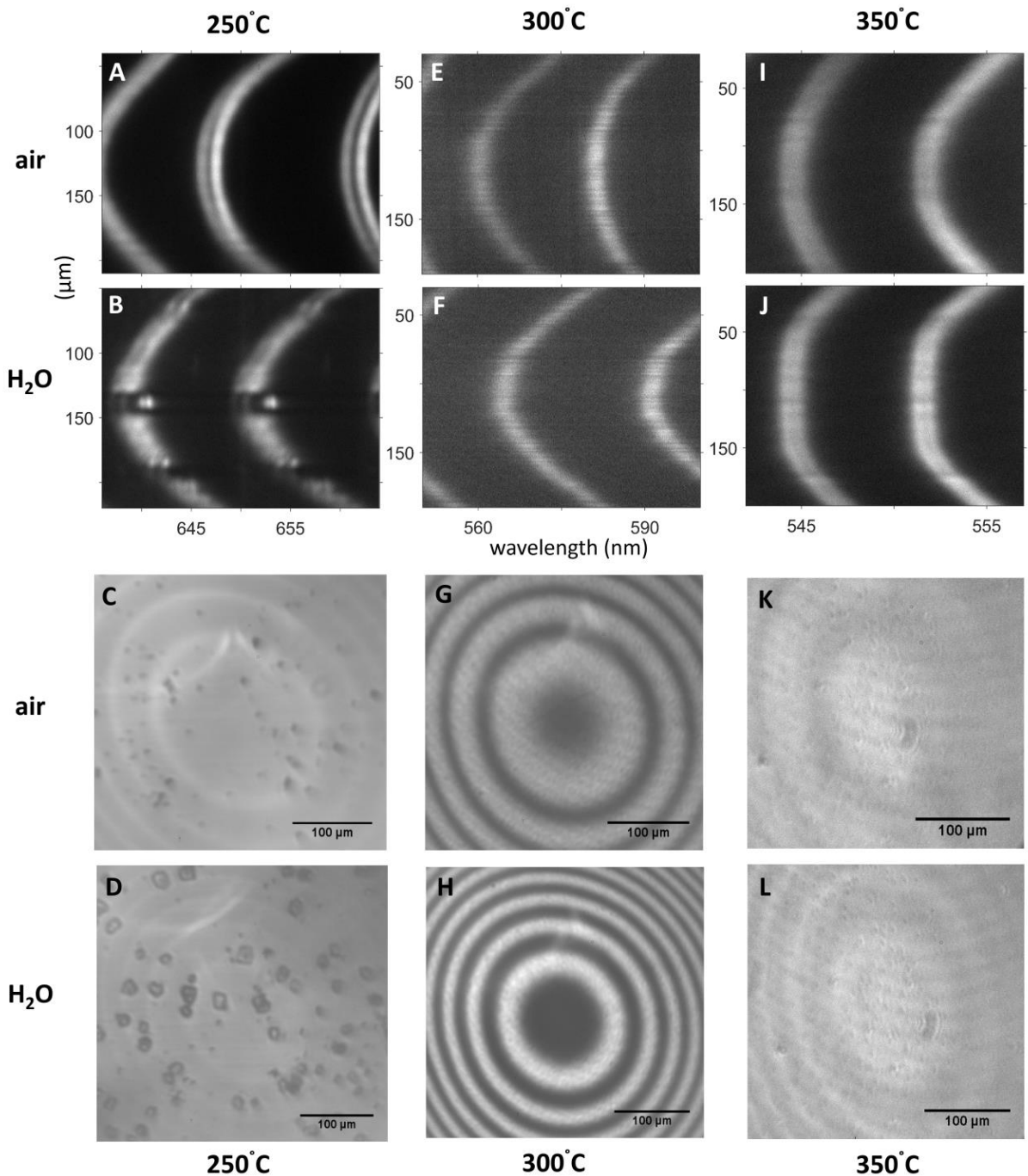

**Figure 5.** *Reactivity of calcite films deposited at 250 °C (CM160511; BAJ1068), 300 °C (CM160714; BAJ1005) and 350 °C (CM770711; BAJ2059) in contact with CaCO$_3$-saturated water (H$_2$O). The two top panels show FECO fringes of calcite and mica surfaces in contact in air, and after the injection of H$_2$O into the SFA chamber. The FECO pattern indicates contact geometry with lateral resolution of 0.624 µm/pixel. The two bottom panels show the corresponding topography of the contacts (top view), with the scale bar = 100 µm in each picture.*

# Reactivity of Calcite Films During Force Measurements in the SFA in CaCO$_3$-saturated solution

The reactivity and thickness of calcite films was additionally followed throughout the whole duration of the SFA experiments (~25 h): initially, when CaCO$_3$-saturated solution was injected into the SFA chamber while the surfaces were kept in a close contact under a constant load (I), during the force measurements (FR – loading-unloading cycle), and while keeping the surfaces in contact overnight (T; ~10 h) under a constant applied load (the detailed SFA experimental parameters can be found in SM, **Tables S6, S7**). We followed changes in a hardwall position (HP – a smallest separation between the surfaces at a given maximum load) with time in two configurations: in an asymmetric system, with one atomically smooth mica surface against a rough calcite layer **(CM; Figure 6)**, and in a symmetric system, using two rough calcite surfaces **(CC; Figure 7)**. The initial calcite thickness was determined at the beginning of each experiment by bringing two surfaces into a contact in air, at a relatively high load, sufficient to elastically flatten the rough samples in contact. This often led to crushing of the highest asperities but provided a more representative thickness of the calcite layer. The surfaces were then shifted laterally to find a new contact before performing the subsequent measurements. For these, much lower loads were applied, so that µm-sized crystals would be preserved on the surface. Large initial values of HPs were related to the presence of such large crystals on the surface. The shift in HP with time was then used as a measure of the rate of calcite dissolution or growth, although for rough surfaces with a nonuniform distribution of asperities it might not be simply proportional to the dissolved volume of calcite.

As previously observed, the films deposited at 250 °C recrystallized immediately upon the injection of CaCO$_3$-saturated solution, and were later more stable, showing a slow but progressive growth of µm-sized crystals **(Figure 6A)**. The topography of these surfaces significantly differed from the other films due to the initial recrystallization, and even though the absolute separation between the surfaces was of the orders of µm, the actual contact areas could be much higher due to the presence of >5 µm-sized crystals (SM, **Figure S15**). The relatively flat and smooth faces of these large crystals could fully come into contact with mica, unlike the nm-sized surface asperities typical for the more stable films, which could only form small, discrete contacts within the nominal contact area. The films deposited at 300 and 350 °C underwent more limited changes in calcite thickness in both surface configurations. On the whole, major shifts in HP occurred upon equilibration with the injected solution, and during FRs, when the surfaces were constantly brought in and out of contact, with less changes during T (contact overnight). In CC, the surfaces generally dissolved throughout the experiments, but for some experiments could grow after a few hours **(Figure 7E, G)**. The morphology of partially dissolved calcite film after the SFA experiment is shown in **Figure 2D**. In CM, surfaces generally exhibited a mixed behavior, but could sometimes grow initially **(Figures 6A, C, E),** and grow markedly in later stages **(Figures 6B, E).**

Noteworthy, the initial dissolution of the films in contact, upon injection of the $CaCO_3$-saturated solution (I; first 0.5h), occurred in a calcite-calcite (CC) system to a much bigger extent than in calcite-mica (CM) configuration, even if accounted for two calcite layers in CC **(Figures 7D, E).** We do not quantify these dissolution rates precisely due to variations in initial temperatures (22.6 to 25.5 °C in CM, 21.7 to 23.6 °C in CC), applied loads (26–746 mN/m in CM, 18–593 mN/m in CC), contact areas (CC<CM<<0.03 mm$^2$) and resulting pressures (CC>CM>>0.01 MPa; minimum pressure calculated assuming a contact radius of 100 μm, and the minimum applied force of 18 mN/m for smooth surfaces), however, we did observe a major difference in HP shifts between the two surface configurations (SM, **Figures S17, S18**). The HP in CC always decreased within the first 30 min and the HP shift varied between 1 and 244 nm in 11 experiments (with a decrease >50 nm for 7 of them), whereas in CM decrease in HP was evidenced in 2 experiments (between 1 and 13 nm) and an increase in HP for 4 of them (0.3, 5, 6 and 427 nm). The very high increase of 427 nm corresponded to the most reactive calcite films grown at 250 °C (CM160429), the behavior of which was discussed previously. Despite calcite solubility significantly affected by temperature, there was no correlation between either initial temperature or temperature change with the shift in HP (ΔT within initial 0.5 h ranged from 0.02 to 0.20 °C in CM and 0.02 to 0.30 °C in CC). Similarly, no correlation between the shift in HP and the initial pH (9.38–9.89 in CM; 8.46–9.90 in CC) or the initial saturation index (SI) with respect to calcite (-0.48–0.13 in CM; -0.97–0.05 in CC) was found. Three possibilities for less dissolution in CM are proposed: 1) dissolution of calcite was influenced by the separation between the surfaces and topographies of contacts; 2) the mica surface affected dissolution/growth of calcite; or 3) dissolution of calcite in CC was enhanced by a higher pressure acting on asperities than in CM.

Additionally, in both surface configurations, major recrystallization events were sometimes observed after several hours, when the surfaces were kept constantly in contact (during T) under moderate applied loads (<200 mN/m). In such events, the crystals growing in or near the contact region were found to push the opposing surface away (**Figures 6B, E; Figures 7E, G**). These events were not correlated with an increase of temperature in the SFA, and therefore were not caused by the decreasing calcite solubility with increasing temperature. Interestingly, such extensive recrystallization of the calcite layer took place only in the case of the least rough contacts, where real contact areas were the largest, and the surfaces could be approached very close to each other (121 nm > initial HPs > -7 nm). In general, the solutions used in the experiments were slightly undersaturated with respect to calcite and represented values for systems equilibrated with calcite under low $CO_2$ partial pressures (SM, **Table S7**). Therefore, any significant growth of the calcite layers was not expected. Because all the major recrystallization events were preceded by a small decrease in calcite thickness (HP shifts between -8 and -47 nm), we suggest that the subsequent rapid roughening of the calcite layer was triggered by an increase in supersaturation in the contact region, due to the prolonged calcite layer dissolution and the limited diffusion from the contact region to the bulk. Additionally, after the recrystallization events, during the further FRs, such recrystallized layer could be deformed as indicated by progressive decrease of HP in the consecutive FRs **(Figure 6E, Figure 7G)**. Such deformation could be related to the smashing of roughened or loose crystals, some horizontal creep in the contact position, or to plastic deformation of calcite. The horizontal creep in the contact position was evidenced from the FECO after the recrystallization only in the case of CM170105, during which the FECO fringes moved laterally (after ~12h; **Figure 7E)**. In the CM system, where adhesive forces between the surfaces were measured, recrystallization of smoother contacts **(Figure 6A, B, D, E)** sometimes led to a significant roughening, decrease in a contact area and a resulting absence of adhesion **(Figure 6E)**. In such cases, the surfaces separated by growing crystals showed a small jump-out followed by the surfaces drifting out of the contact **(Figure 6E)**. Less frequently, rougher mica-calcite contacts evolved into smoother, adhesive contacts by a ripening growth of some asperities and a resulting increase in the real contact area between mica and calcite surfaces **(Figure 6D)**. The morphology of a calcite film that underwent a significant recrystallization is shown in **Figure 2C**, showing an area with coarser and more irregular crystals. Such coarsening was evidenced only in some regions of the samples and it was not possible to distinguish where the contact regions used for the SFA measurements were.

### Effect of Contact Topography

We observed that in the initial experimental stages, calcite in CC dissolved faster when the distance between the surfaces was larger (SM, **Figure S17**), whereas the calcite in CM grew or slightly dissolved, showing no clear dependence on the surface separation. In general, for two rough CC surfaces, the fluid volume between the surfaces should be larger than for CM, due to both the larger gap thickness and the larger area between the asperities. Larger asperities simultaneously increase the volume and distance between the two surfaces and decrease the actual area of contact between the surfaces. However, large actual areas of contact can sometimes be formed by very big asperities (e.g. calcite grown at 250 °C; CM160429; **Figure 6A**). In CM, the contact areas should be larger and gaps narrower, since the mica surface is atomically smooth. We observed a dependence of the HP shift on the initial gap size in CC, for the dissolving calcite surfaces with initial HPs <0.6 µm (9 out of 11 experiments; SM, **Figure S17**). This correlation was not present for the CC and CM surfaces with much bigger (>1 µm) asperities, for which the volume loss on dissolution corresponding to a given shift of HP was much bigger than for smaller asperities. Therefore, the faster calcite dissolution in CC and growth in CM could be solely related to the contact geometries, which possibly limited mass transport and enhanced saturation of the confined fluid in the gap. It has been recently shown that the thicknesses of boundary layers, which formed close to the surfaces of various dissolving carbonates, depend on the carbonate dissolution rate, kinetics of precipitation of other phases, and diffusion rates of the ionic species into the boundary layer, and can be of the order of µm [90]. Here, in the case of rapidly dissolving 250 °C films, the whole volume of the solution in the gap between the surfaces could have become quickly supersaturated, even for the wide µm-sized gaps. In contrast, the more stable calcite films (300, 350 °C) displayed much lower dissolution rates that were possibly of a comparable order as the diffusion rates out of the gap. Therefore, the boundary layer thicknesses were relatively thinner, and were probably controlled by the diffusion rates (and the contact geometries) to a greater extent than in the case of the very rapid dissolution rates for the 250 °C films. This would explain why the measured dissolution rate was the fastest for the roughest 300 and 350 °C surfaces in CC. Previous experimental studies of pressure solution between mineral surfaces in contact have pointed to some dependence of dissolution processes on mass transport across the contacts and/or measured a dissolution rate varying with time in a complex way [92-98]. For example, the dissolution of dissimilar mica and quartz surfaces in contact, followed in the SFA, have been found to gradually slow down, despite the undersaturation of the bulk solution [92], which could suggest a developing chemical equilibrium with the solution in the gap. Later, however, this phenomenon has been explained as being due to the reprecipitating silica gel, sealing the contact junction and limiting the diffusion [93], or due to an electrical potential difference between these two dissimilar surfaces [99-100], which all points to the complexity of solid dissolution at contacts. Roughness can additionally influence the dissolution process: rough contacts can be depicted as sets of small pores between the contacting asperities, with a limited connectivity and transport between them, especially for calcite polycrystalline films with nm to µm-sized grains. It is reasonable that mass transport along such contact occurs via different mechanisms depending on the pore size and their connectivity [101], and is additionally affected by an inhomogeneous stress distribution [102]. Recent models of a gap distribution in contacts between a randomly rough surface against a smooth, elastic surface predicted an existence of contact and non-contact patches that displayed different fluid percolation properties [103]. Even in a stationary fluid, like in the SFA, this might affect the mass transport through the reactive contacts (of nominal areas <0.3 mm$^2$), with regions more and less isolated from the bulk solution. As such, the dissolution of calcite should be more limited for less rough contacts in CC, and especially in CM, in which cases 'sealing' of some void regions of the contact should be more effective, and saturation with respect to calcite should be achieved faster, due to the generally smaller volumes of these voids. We found such dependence for most of the experimental data (SM, **Figure S18**). Additionally, the recent experiments with single calcite crystals growing against a glass surface have revealed a presence of large cavities in the contact region, in which calcite could be in equilibrium with a trapped liquid over a long time, despite a separating water film as thick as ~50 nm [104].

The limited mass transport along the contacts can also explain the major recrystallization events, in which the calcite layer grew rapidly after a prolonged period of dissolution, and pushed the opposing surface away on growing, despite the applied load. In such cases, the fluid supersaturation reached a sufficient level for the nucleation to occur in the gap, and the force of crystallization associated with the growing crystals [105] could overcome the normal compressive stress applied to the confining surface.

### Effect of Dissimilar Surfaces

Dissolution of calcite could have been additionally affected by a proximity of the mica surface. We did not see any direct influence of the mica surface chemistry on the recrystallization of the 250 °C films, however, given the very high reactivity of these surfaces, possible effects could be negligible. Our measurements with the 300 and 350 °C films, revealed only a limited initial (first 0.5 h) dissolution and sometimes a growth of the calcite layer in CM, the latter never initially observed in CC (SM, **Figure S17**). The mica basal surface is negatively charged ($\sigma$ = ~-0.3 C/m$^2$) and the near-surface distribution of the counterions, such as $Ca^{2+}$, can be many times higher than in the bulk. Higher valency ions preferentially adsorb on the surfaces [16], and their exchange with other species from the bulk is very slow [92, 106]. The initially dissolving calcite surface could be a source of $Ca^{2+}$, which upon binding to mica might have additionally contributed to the supersaturated conditions in the gap (the used solutions were slightly undersaturated; SM, **Table S7**). Interestingly, Å-size $CaCO_3$ crystals has been reported to grow between mica surfaces separated by <1 nm, even in undersaturated bulk conditions, and can occur in natural systems where calcite has been observed to grow between mica layers [106]. This suggests that mica can promote the growth of calcite, in contrast to silica enhancing the dissolution of halite [107] or mica enhancing the dissolution of quartz [108]. Dissolution in the latter system has lately been attributed to electrochemical corrosion interactions in the electrical double layers of these dissimilar surfaces [99-100]. In our experiments, there was definitely a difference in surface potentials ($\psi_0$) of mica and calcite in $CaCO_3$-saturated solutions, with mica being more negatively charged (~-70 mV; SM, **Figure S13**) than calcite (~-20 mV, [36, 69]). If the recrystallization of calcite in the proximity of mica could be triggered by an overlap of the electric double layers of these two minerals (resulting in a transfer of counterions, altered surface potentials on both sides and a lowered energy barrier for dissolution [100]), it would rather be expected that the dissolution of calcite (higher $\psi_0$) was enhanced by mica (lower $\psi_0$). In such case a transfer of $Ca^{2+}$ counterions would take place from calcite to the mica surface. It could be then possible that the observed growth corresponded to calcite precipitating on the mica surface. This possibility could not be ruled out in FECO (not possible to distinguish on which surface the process took place), nor with SEM after the SFA experiment (crystallites could be too small). However, as will be discussed later, we observed a strong attractive interaction in CM (also in case of the growing calcite surface), which was never present in CC [35]. This supports the suggestion that calcite did not nucleate on mica, at least not in the contact region, as this would have led to an absence of attraction. If mica does enhance the dissolution of calcite, then the observed growth had to be caused by the resulting supersaturation conditions in the gap between the surfaces. The growth of the calcite layer was rather limited (several nm/h, apart from one experiment with the most reactive calcite film deposited at 250 °C), which could suggest that the observed growth was a result of dissolution/precipitation in the gap [109].

### Effect of Applied Load

Two pressure driven deformation mechanisms may be operating in these experiments: 1) enhanced dissolution due to pressure solution and 2) plastic deformation of calcite crystals. Pressure enhances the solubility of the stressed solid, driven by an increase in chemical potential relative to a stress-free surface [110]. In the CC system the total area of contact was much smaller than in the CM system. Thus, if the same load was applied in CM and CC and the distribution of asperities was similar (density, sizes and height distribution), stress transmitted at these discrete contacts with a smaller overall area in CC would be much higher [111] than in CM. We would therefore expect to see more dissolution in the CC system if dissolution was mainly stress-induced. Neither CM nor CC systems displayed any clear correlation between the maximum applied load and the shift of HP, However, as there was no way of estimating the real contact areas precisely, the shift in HP could potentially have been related to the applied pressure. In general, the real contact areas should be the smallest in the experiments with the roughest surfaces, resulting in the highest pressures acting on asperities. Thus, if pressure solution was causing the decrease in HP, this decrease should be most pronounced for the roughest calcite surfaces. This is not what we observed. When normalizing by the maximum applied load, the shift in HP was not correlated with the estimated surface roughness. This suggests that the dissolution of calcite was not primarily driven by pressure solution. Previous studies, although at lower effective stress, have also shown that the dissolution rates of the stressed calcite surfaces in contact with water were not correlated with the applied stress magnitude [88, 112]. Also, calcite subjected to a normal load by an indenter in solutions, has displayed the strain rates that were not clearly dependent on the applied stress [29].

Any possible plastic deformation of the calcite surfaces would be expected to take place when the surfaces were first pushed into contact [16]. The decrease in thickness progressing with time, observed also during unloading **(Figures 11, 13B),** cannot be thus explained by plastic yielding. However, it was highly likely that plastic deformation of the highest asperities took place on the initial loading, leading to an irreversible deformation and flattening of these asperities. With the used range of loads (>850 mN/m), the pressure at the highest asperities most probably exceeded 1 GPa (at which plastic deformation of single calcite crystals has been observed [113]). It has been observed that the loading of salt crystals usually causes an initial plastic deformation of the surface asperities subject to the highest stress values [94]. Nevertheless, even though the plastic deformation was likely to occur on initial loading, it cannot explain the variation in rates of the HP decrease with time for the different experiments.

## Reactivity of Calcite Films During Force Measurements in the SFA in MEG

In one set of experiments, we additionally followed the evolution of HP in a CC system in MEG **(Figures 7A, B)**. This system is of interest due to the observed mechanical strengthening of carbonate rocks saturated with MEG [15], and the previously measured adhesive forces measured between two calcite surfaces in MEG [35]. In the case of low roughness surfaces, the calcite thickness changed only slightly (HP shift of ~0.5 nm) within 25 h **(Figure 7A)**, and no major recrystallization was observed within 3 experimental days with the same set of surfaces. However, for the rougher films, calcite was progressively dissolving, as indicated by a steady decrease in HP during FR (~4 nm/h in day 1 and ~6 nm/h in day 2; **Figure 7B)**, and this dissolution rate was much smaller when these surfaces were kept constantly in contact overnight (1 nm/h; T; **Figure 7B)**. The solubility of calcite in MEG is affected by a relatively high $CO_2$ solubility in this organic solvent. In general, however, the solubility of calcite in MEG is lower than in water (9.2 mmol/kg in 95 wt% of MEG, $P_{CO2}$ = 1 bar, NaCl = 0.5 M, T = 25 °C, in comparison with 14.71 mmol/kg in the same conditions in pure $H_2O$; as measured by Sandengen [114]). Here, because the MEG solutions were not presaturated with respect to $CaCO_3$, some dissolution of calcite films upon equilibration with MEG was expected. Thus, the observed minor changes in calcite thickness in the case of a very smooth contact, and the limited dissolution of rougher surfaces when kept in a close contact (T), could be again attributed to a higher saturation in the gap that arose due to the more limited mass transport across the contacts in these cases. As diffusion rates in MEG are lower than in water because of the high viscosity of MEG (16.9 cP at 25 °C), these confinement effects were more enhanced in comparison with the $CaCO_3$-saturated solutions. Perhaps, due to more limited solubility and prolonged induction times for $CaCO_3$ precipitation in MEG [115], we never observed any major recrystallization of calcite films, even with the smoothest surfaces (CC170704) over 3 days.

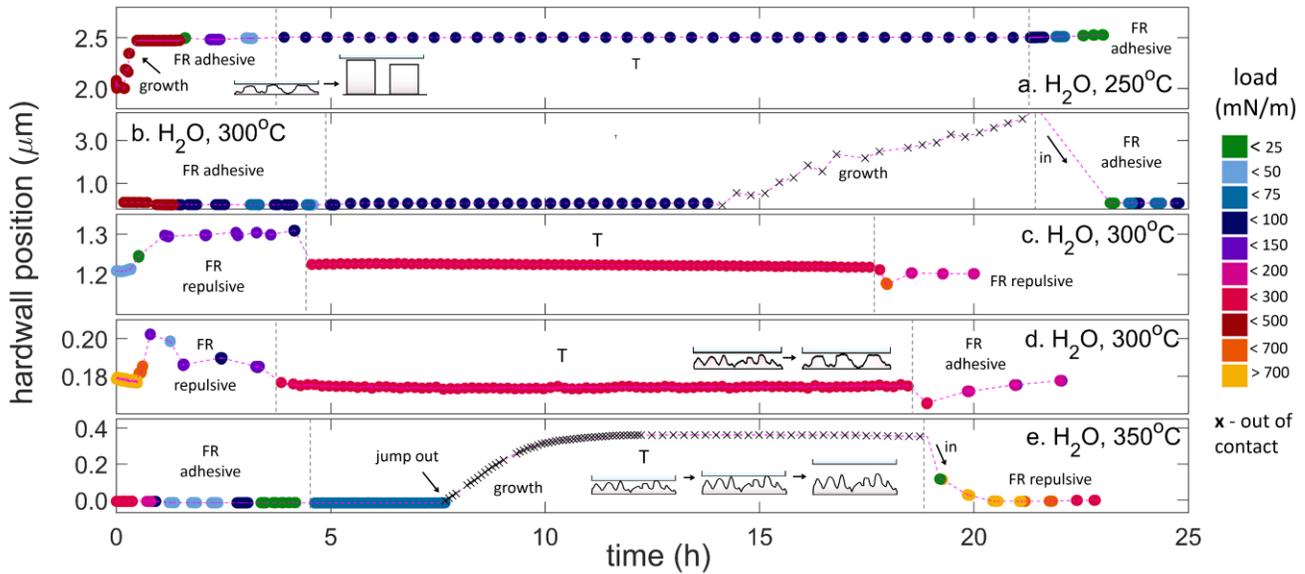

**Figure 6.** *Evolution of hardwall position (HP) with time in SFA experiments between mica and calcite surfaces in CaCO$_3$-saturated solution (H$_2$O). Major recrystallization events were indicated in the figure, where the symbol x refers to surfaces separating from contact due to calcite growth. FR – HP measured during loading-unloading cycles with an indication of adhesive or repulsive interaction between the surfaces; T – HP measured while keeping surfaces in contact under constant load. The used calcite surfaces were deposited at different temperatures: a. 250 °C (CM160429; BAJ1068), b. 300 °C (CM160714; BAJ1005), c. 300 °C (CM170309; BAJ2046), d. 300 °C (CM170222; BAJ2046), and e. 350 °C (CM170711; BAJ2059). Note big differences in y axes scales.*

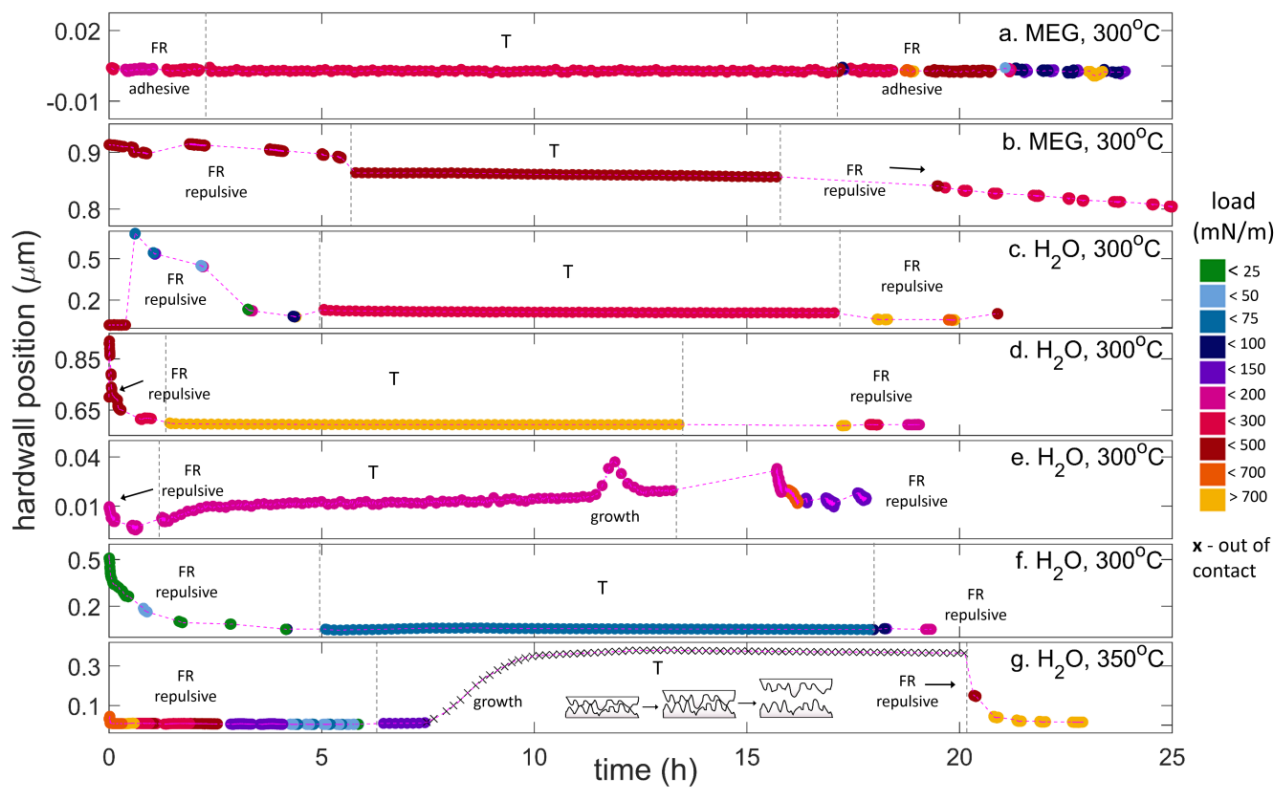

**Figure 7.** *Evolution of hardwall position (HP) with time in SFA experiments between two calcite surfaces in ethylene glycol (MEG) or in $CaCO_3$-saturated solution ($H_2O$). FR – HP measured during loading-unloading cycles; T – HP measured while keeping surfaces in contact under constant load. The used calcite surfaces were deposited at: a. 300 °C (CC170704; BAJ1041), b. 300 °C (CC170912; BAJ2057), c. 300 °C (CC161101; BAJ1025), d. 300 °C (CC161109; BAJ1026), e. 300 °C (CM170105; BAJ2046), f. 300 °C (CM170410; BAJ1041), and g. 350 °C (CM170713; BAJ2057).*

# Force Measurements Between Calcite and Mica Surfaces

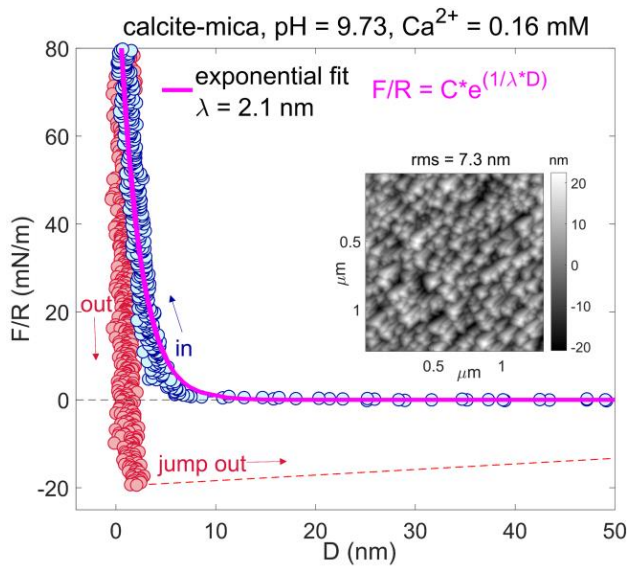

**Figure 8.** *Representative data from SFA measurement of the force (normalized by contact radius, F/R) between calcite and mica surfaces as a function of separation (D), showing an attractive interaction in a CaCO$_3$-saturated solution. The D values are shifted so that HP = 0 nm at the maximum applied load. The green line shows an exponential fit to the FR on approach, F/R = C·exp(1/λ·D), used to estimate a roughness magnitude, where λ is an exponential decay length, proportional to the average local roughness in the contact region. The inset shows an AFM height map of the calcite film before the experiment (CM160711, BAJ1005, rms = 7.3 nm; scan size 1.2x1.2 μm).*

Forces in an asymmetric setup, between calcite and mica surfaces (CM), were measured with the SFA in CaCO$_3$-saturated solutions (SM, **Table S7**). Both adhesive and repulsive forces were observed in this system, depending on the calcite surface topography. The attractive forces could be measured whenever a sufficiently large contact area was established between the surfaces, which usually corresponded to the least rough calcite surfaces. Additionally, attractive forces were observed with the calcite surfaces grown at all three temperatures 250, 300 and 350 °C, showing no observable effect of the deposition temperature, and thus the resulting film characteristics on the type of forces. A representative attractive FR in CM is shown in **Figure 8:** forces on approach were in most cases purely repulsive, with no resolved jumps-in, followed by relatively large adhesive jumps-out on separation. The magnitude of pull-off forces varied between consecutive FRs in each experiment, suggesting an evolution of contact topographies with time. Adhesion was present both when the calcite layer was growing (CM160429, CM170222) and dissolving (CM160714, CM170711).

We first investigated how the measured pull-off forces corresponded to the changing surface topography of calcite within 2 consecutive days for each experiment **(Figure 9, 7, 8)**. Since both magnitude and onset of attractive and repulsive forces will depend on surface roughness [43, 54, 116-118], we used an exponential decay length $\lambda$ of the approach part of the repulsive FRs to semi-quantify the changes in surface roughness, according to F/R = C·exp(1/$\lambda$·D) **(Figure 8)**. We assume that in our system, the major contribution to the measured repulsive forces is the energy needed to compress multiple surface asperities elastically, which can be represented with a Hertzian-type deformation [16, 45]. As such, $\lambda$ should be proportional to the average local roughness of the contact areas (involving only the asperities 'felt' by the approaching surface upon loading), but not to the overall rms roughness of the samples [119]. Even though the measured FRs did not always show a purely exponential behavior on approach (which is typical for rough surfaces with a random, Gaussian-like distribution of heights [45, 116]), $\lambda$ was still a satisfactory parameter to account for the magnitude of changes. We expect the electric double layer force in our system to have a minor contribution to the measured repulsion (EDL in a $CaCO_3$-saturated solution is long-ranged, with the onset at ~70 nm, and of very low magnitude; $EDL_{max}$ <6 mN/m for smooth calcite-mica surfaces; see Eq. 4). The repulsion could be affected by the hydration force, as discussed later.

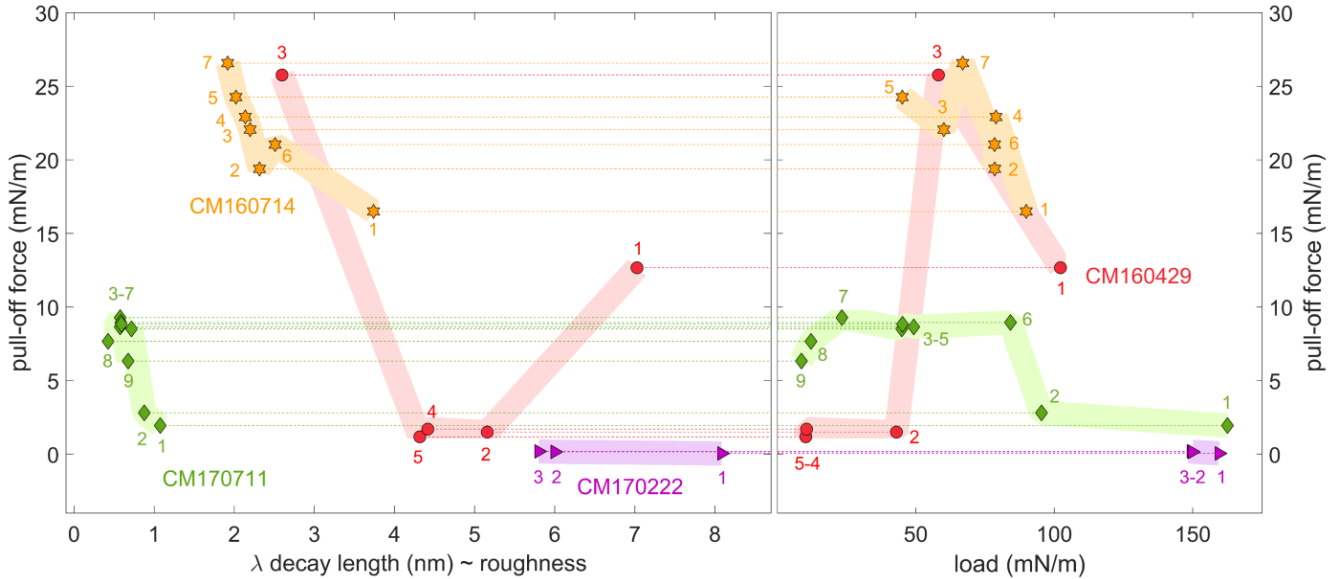

**Figure 9.** *Pull-off forces (minimum of measured attractive force upon separation) as a function of estimated roughness magnitude (λ – exponential decay length; left), and as a function of the maximum applied load (right) between calcite and mica in $CaCO_3$-saturated solution. The numbers represent the order of FR measurements. Each color corresponds to a different set of calcite-mica surfaces (CM160429 – BAJ1068, 250 °C, runs 1,2 on day 1, runs 3–5 on day 2; CM160714 – BAJ1005, 300 °C runs 1–5 on day 1, runs 6,7 on day 2; CM170222 – BAJ2046, adhesive on day 2, 300 °C, CM170711 – BAJ2059, 350 °C, adhesive on day 1). Horizontal lines connect corresponding experimental points. The pull-off forces (and loads) for CC170222 were 1) 0.05 (159.5), 2) 0.16 (149.9), and 3) 0.19 (149.4) mN/m. The absolute magnitudes of forces in different experiments are not comparable due to major differences in contact topographies and areas.*

In general, adhesive forces in CM were detected for FRs with the decay lengths λ on approach <8.5 nm **(Figure 9; 16)**. The dependence of both the estimated roughness (λ), and the maximum applied load on the pull-off forces in CM is shown in **Figure 9**. In all cases the pull-off forces increased with decreasing λ, and thus with smaller roughness of calcite in contact regions. This dependence was not apparent for CM160429 (deposited at 250 °C), however, the topography of these recrystallized samples differed a lot (>5 µm-sized crystals**;** sketch in **Figure 6A**) in comparison with the 300 and 350 °C films (nm to µm-sized asperities). Thus, the resulting real contact areas for CM160429 could be much bigger, and the measured pull-off forces were as a consequence more dependent on the applied load **(Figure 9)**. In this experiment, much weaker adhesive forces (runs 2, 4, 5) clearly corresponded to smaller applied load values (<50 mN/m). There was no correlation between the maximum applied load and the magnitude of pull-off forces for the other experiments (CM160714, CM170222, CM170711).

Additionally, we considered how the forces varied with the experimental time **(Figure 9)**. Interestingly, the pull-off forces became stronger in consecutive FRs for all experiments, while the calcite surfaces were progressively dissolving (CM160714, CM170711; **Figure 6B, E**), or progressively growing (CM170222 and CM160429; **Figure 6D, A**). This suggests that the adhesive forces were affected by the changes in contact topographies, which could become less rough with time during both calcite growth and dissolution. Consequently, the contact areas and the pull-off forces increased. It was not clear if the decrease in contact roughness with time was directly related to surface reactivity (dissolution/growth) or related to another type of surface deformation, which continued with time during the repeated FRs, independently of dissolution and growth. We also observed an initial decrease in roughness (λ) during FR in a repulsive CC system, before calcite surfaces underwent recrystallization (e.g. **Figure S21**).

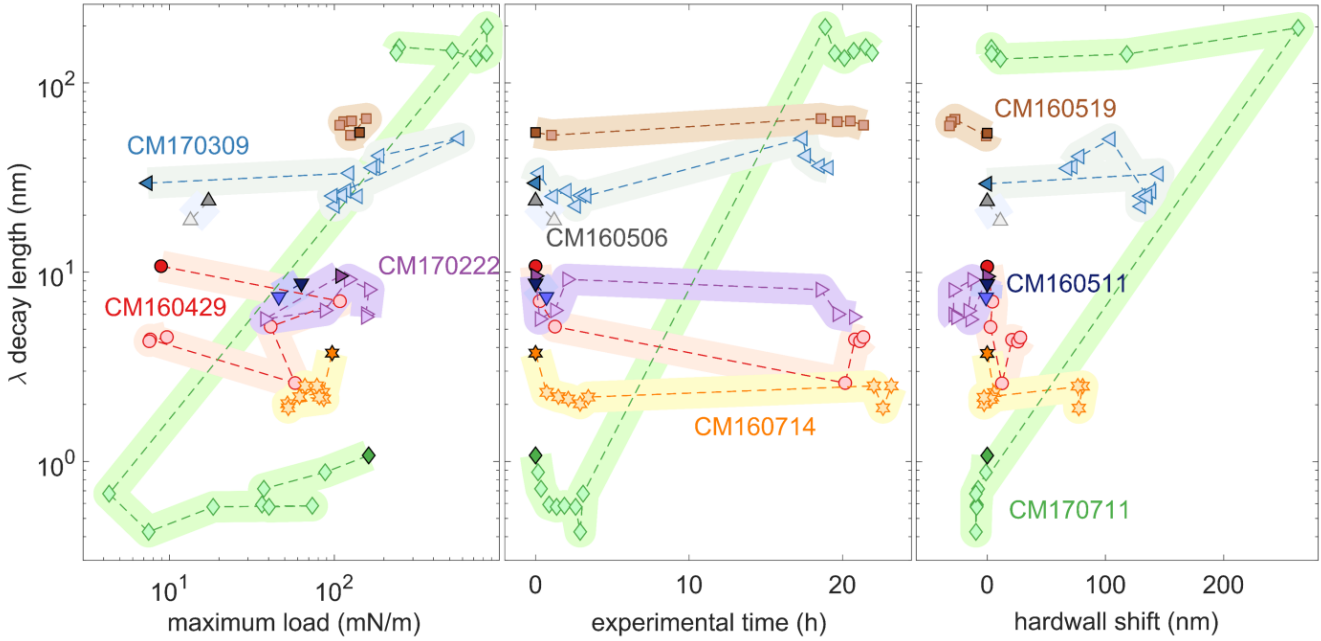

**Figure 10.** *Exponential decay lengths λ (~ roughness) of the FR curves measured with the SFA on approach between calcite and mica surfaces, as a function of the maximum applied load, experimental time, and a shift of HP. The shown data includes both repulsive and adhesive FRs. The dashed lines connect the data points in the order of the measurement and the first measured FR curve in each experiment was marked with a darker-colored symbol. The shift in HP position was given with respect to the first experimental point and was measured at the same applied load value for each experiment.*

We further analyzed changes in roughness (λ) as a function of the maximum applied load, experimental time, and HP shift (relative to minimum HP in the first FR), for all attractive and repulsive experiments in CM **(Figure 10)** to examine what factors had the largest influence on topography evolution. As previously observed, there was no clear relationship between the maximum applied load and roughness (λ) for any of the experiments, suggesting that the magnitude of the applied load was not the primary driver for changes in topography **(Figure 10)**. On the other hand, the evolution of λ was correlated with the total experimental time. In general, within 25 h we measured a major increase of λ (>2 orders of magnitude) only in the case of CM170711, in which experiment a major recrystallization of the surfaces occurred overnight **(Figure 6E)**. In all cases, we observed an initial decrease of λ within the first 5 experimental hours. We did not have information about the roughness evolution while the surfaces were kept in contact overnight (5–20 h), but the first measured λ on day 2 increased relative to the last experimental point on day 1 in most experiments (apart from CM160429 and CM170222). Interestingly, in most cases λ gradually decreased again once the FRs were restarted the following day (20–25 h, except for CM160429 and CM160714). To test whether such periods of decrease in λ with time were related to dissolution or growth of the surfaces, λ was plotted as a function of the shift in HP **(Figure 10)**. Overall, there was no apparent correlation between these two parameters. However, at times when λ was progressively getting smaller in consecutive FRs, we mostly observed a corresponding gradual decrease in HP (CM170711, CM170309, CM160714, CM160519) or less frequently an increase in HP (CM160429, CM170222). The magnitude of these HP shifts rarely exceeded 100 nm. This shows that during FRs, the contact topographies could become smoother both in case of dissolution and growth of calcite layer, and may suggest that not directly the surface reactivity, but the repeated loading of the surfaces could significantly contribute to gradual decrease in roughness at the contacts. This is further supported by the fact that during periods when the surfaces were constantly kept in contact overnight (T) and no loading-unloading cycles were performed, the roughness increased in most experiments.

### Effect of Reactivity and Applied Load on Pull-off Forces

In summary, we found that the magnitude of the adhesive force in CM was related to changes in contact topography with time, which was most likely caused by growth/dissolution of calcite and/or plastic deformation of the highest asperities on the repeated loading-unloading cycles. We suggest that the observed increase in pull-off forces was related to gradual increase of the real contact areas between the surfaces during FRs. It was not clear if the reactivity of calcite surfaces was likely to cause this increase directly, since we observed larger pull-off forces both when calcite was dissolving or growing. It was also possible that contact areas increased due to progressive plastic flattening of asperities on repeated loading. Although the magnitude of the pull-off force rarely depended on the maximum applied load, it could be sufficient to apply any load in the range that we used to gradually flatten the asperities during the FRs.

Perhaps the growth and dissolution of calcite surfaces could alone lead to the increase in contact areas. As previously discussed, the changes in chemical equilibrium in the gap between the surfaces seemed to have the dominant contribution to topography evolution throughout the experiments. It can be imagined that progressive dissolution of the highest asperity tips makes their area gradually larger, leading to higher contact areas in consecutive FRs. Also, the growth of relatively big µm-sized asperities with time could make the real contact areas larger as well. On the other hand, the changes in a surface free energy make the smaller crystals generally dissolve preferentially to the bigger ones [78], and thus the high asperities should not dissolve first. Additionally, the increase of roughness observed when the loading-unloading cycles were stopped and the surfaces were kept in contact overnight, strongly suggests that the mechanical stress during loading in consecutive FRs played a role in increasing the contact areas. The applied load could affect the pull-off forces in two ways: (1) increase the real area of contact reversibly by elastic deformation; (2) influence the real area of contact irreversibly by pressure solution and plastic deformation.

Firstly, in general, for multi-asperity rough elastic surfaces, higher applied loads yield proportionally larger real contact areas [120-122], likely to result in higher adhesion energies and thus higher pull-off forces. However, the pull-off force is also affected by roughness in another way: elastic energy on compressing the asperities is released during the unloading and may be large enough to overcome the adhesive bonds [118], sometimes counteracting the expected increase of the adhesion force due to a higher real contact area [119]. Although varying the applied load in CM170711 between 25 and 100 mN/m (runs 3–7; **Figure 9**) did not seem to have any effect on the measured pull-off forces, this may be because the range of stresses was insufficient to cause a significant increase in contact area for these surfaces, whereas it was sufficient for CM160429, which did display an effect of applied load (runs 1 and 2 on day 1, runs 3–5 on day 2; **Figure 9**).

Secondly, applied load could induce irreversible changes in contact topographies. Whereas in our system an increase in HP was generally related to calcite growth, a decrease in HP could be related to a deformation, which can result from dissolution and/or compression. As follows from the earlier discussion, pressure solution favors dissolution of the most stressed, highest asperities in contact over the stress-free non-contact regions, because of the differences in their chemical potential. Even in the cases where we measured an overall growth of calcite layer, material could be redistributed so that the calcite surface became smoother due to the action of pressure solution. Similar effects would be observed in the case of plastic deformation on the repeated loading. We could not distinguish between these two pressure-driven mechanisms unambiguously for the adhesive FRs in CM. However, using the plasticity index (PI) parameter [117, 123], it is possible to estimate if an elastic or a plastic mode of deformation of the rough surfaces is dominant. PI is based on hardness, elasticity and roughness parameters for a given surface, and is calculated as: $PI = \frac{E_r}{H}\sqrt{\frac{\sigma_h}{r_h}}$, where $E_r$ is the reduced Young's modulus (see Eq. 6), $H$ is hardness, $\sigma_h$ is standard deviation of surface heights as measured with the AFM, and $r_h$ is an average asperity tip radius as estimated from the AFM topography scans. PI was calculated for the calcite films using $H$ = 1.49 GPa [124], and Young's modulus of 72.35 GPa. For one of the smoothest calcite surfaces (BAJ2059; $\sigma_h$ = 4.3 nm, $r_h$ = 55 nm, scan size = 10x10 µm$^2$) we found PI = 7.4, and for one of the roughest surfaces (BAJ2046, $\sigma_h$ = 114 nm, $r_h$ = 520 nm, scan size = 10x10 µm$^2$) PI = 12.5. Such large values of PI >1 suggest that asperities in the calcite films were very likely to undergo plastic deformation, both for very smooth and very rough surfaces.

As such, if we assume that the effect of pressure is to smoothen the surface [110, 125-126], and the effect of chemical dissolution is to roughen the surface [37, 40, 84, 88, 112, 127], we would expect pressure to influence the topography, as we see decrease of λ with time. Even though the magnitude of pressure did not correlate with the shifts in HP, higher pressure concentration at asperities could still influence which regions dissolve preferentially. However, as the calcite films were prone to plastic deformation, we believe that the contact areas most likely increased due to progressive flattening of the highest asperities on repeated loading-unloading cycles. Nevertheless, in all cases we saw an increase of the measured pull-off forces with time related to a topography evolution towards higher contact areas, which was interrupted by the periods of time when surfaces were kept in contact for several hours and usually became rougher.

### Attractive Interactions in CM

It is interesting that we measured such strong pull-off forces in CM. From comparing the theoretical VdW interaction in CC, in CM and between two mica (MM) surfaces, VdW attraction should be slightly higher in CC and almost the same in CM and MM (Hamaker constants of 1.44, 1.35 and $1.34 \cdot 10^{-20}$ J, respectively [67]). We measured weaker pull-off forces in CM than in MM (SM, **Figure S14**) in $CaCO_3$-saturated solutions, which is most likely due to calcite roughness and much smaller contact areas in CM. The average pull-off force in MM was 85 mN/m (varied between 61 and 11 mN/m in 6 experiments; data not shown), whereas in CM it ranged from 0.1 to 26 mN/m, depending on the roughness. The strongest adhesive force in CM was thus only ~3 times smaller than in MM. Since we assume that even for the smoothest calcite surfaces the real contact area was only a very small fraction of a nominal area in the SFA, we would expect the adhesive force and adhesion energy to be even weaker in CM, if the attraction was only related to VdW interaction. Therefore, it is likely that other non-DLVO attractive effects were also present. According to zeta potential measurements [36], at high ion concentrations (that could be present due to dissolution of calcite and a higher concentration in the gap between the surfaces) the calcite surface can become positively charged, while the mica surface is still negatively charged, leading to a longer-ranged electrostatic attraction of the double layers. Additional electrostatic effects, such as ion correlation forces, could also enhance the adhesion in CM. These interactions are particularly strong in presence of higher valency cations (e.g. $Ca^{2+}$) even in dilute solutions, and are possibly responsible for limited swelling of $Ca^{2+}$-clays [16, 128-130].

# Force Measurements Between Two Calcite Surfaces in CaCO$_3$-saturated Solutions

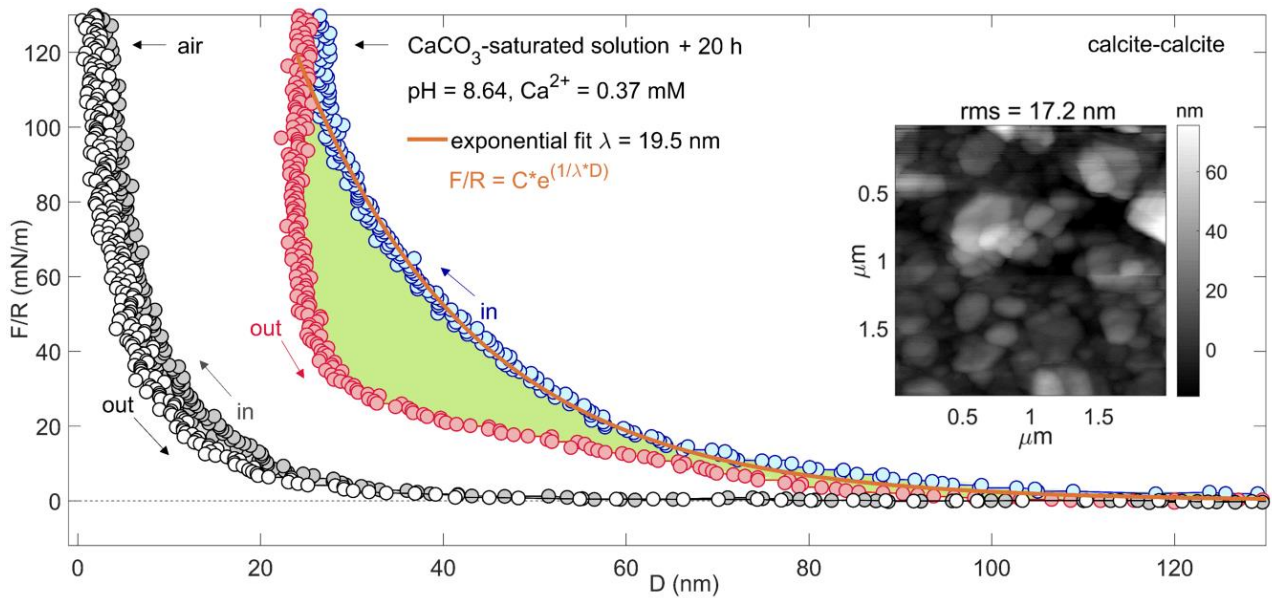

*Figure 11.* Representative SFA force-distance curves between two rough calcite surfaces in air and in CaCO$_3$-saturated solution. The D values were shifted so that HP = 0 at the maximum applied load in the FR measured in air with no prior drying with N$_2$. The shown FR in CaCO$_3$-saturated solution was measured 20 h after the FR in air in the same contact region. The green area marks the hysteresis between approach and separation. The inset shows an AFM height map of the calcite film before the experiment (CM170501, BAJ2046, rms =17.2 nm; scan size 1.8x1.8 µm$^2$).

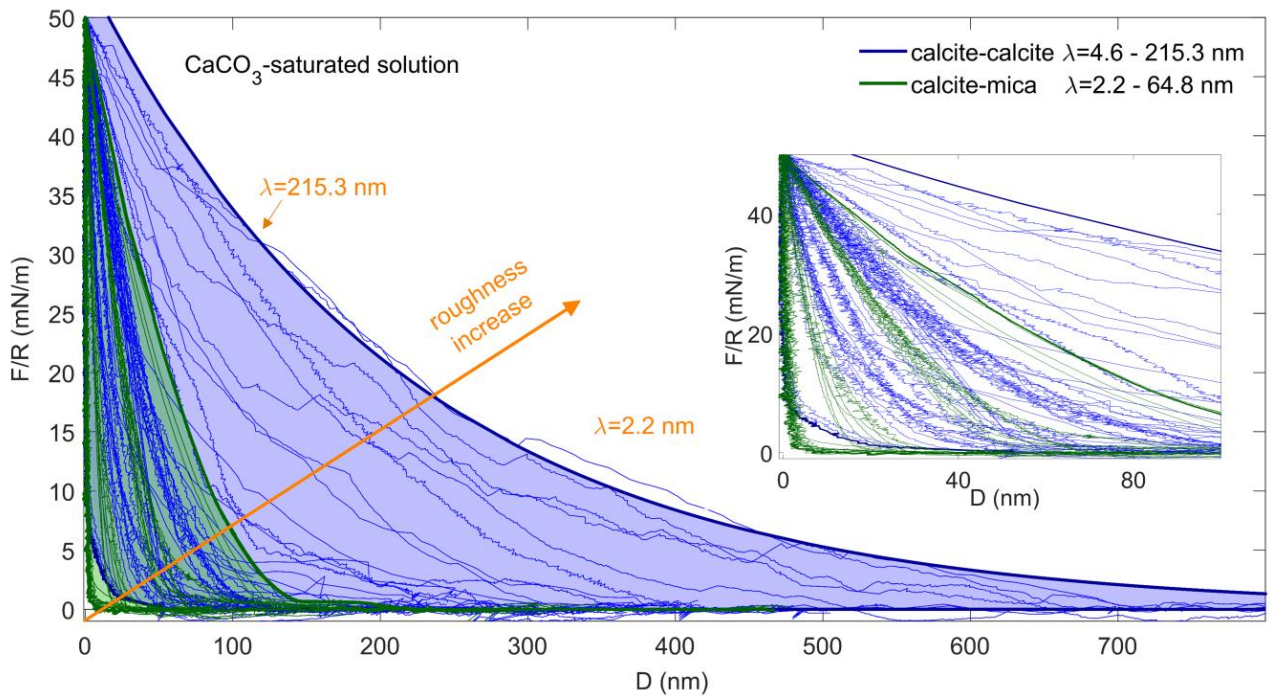

**Figure 12.** *Repulsive forces as a function of a distance D between two calcite surfaces (11 different sets of surfaces; blue) and between calcite and mica (7 different sets of surfaces; green) in $CaCO_3$-saturated solution. The magnitude and onset of the repulsive forces was strongly influenced by the roughness of calcite films, with exponential decay length λ ranging from 2.2 to 215.3 nm. Only the FRs on approach are shown. The D values were shifted so that D = 0 nm was located at the applied load of ~50 mN/m for all the FRs. The shadings mark an approximate range of the measured forces. The inset zooms the D region between 0 and 100 nm.*

Forces in a symmetric setup, between two rough calcite surfaces (CC), were measured with the SFA in $CaCO_3$-saturated solutions (SM, **Table S7**). Only repulsive forces were observed in this system, in solution as well as in air. A representative FR in CC is shown in **Figure 11**. Whereas the FR in air showed only a small hysteresis between loading and unloading, we observed significant hysteresis in the FR measured in the solution. Such hystereses, fully located in the repulsive region of the force, were measured for all the experiments in CC (and in CM with no adhesion force) and will be discussed later. The representative FRs in air and in solution **(Figure 11)** displayed very different magnitudes and onsets, along with a shift of HP related to calcite growth (after 20 h) in solution. These differences indicated progressive changes in contact topography of the reactive calcite surfaces. The magnitude and range of the repulsive forces strongly depended on the roughness of the surfaces, or, more precisely, on the roughness in the contact areas in the SFA **(Figure 12)**. Any major roughening of the surfaces during the experiments led to a significant increase in the magnitude and onset of the measured repulsion **(Figures S20, S21)**. The exponential decay lengths of the FR on approach, used to estimate the local roughness in contacts, varied between 5 and 216 nm for two rough surfaces in CC, and between 2 and 65 nm for one rough surface in CM. Even though adhesive forces were measured in CM for the surfaces with a λ <8.5 nm, we did not measure any adhesion even for the smoothest set of surfaces in CC (λ = 4.6 nm). The rms values of the calcite surfaces measured with the AFM after the SFA experiments ranged from 9 to 228 nm **(Figure 4, Table S9)**, however they did not always correlate well with λ for a given set of surfaces in the SFA, perhaps because it was not feasible to image the same region of the sample. No 'jump-in' discontinuities [36] in the repulsive forces on approach were detected, presumably due to the high roughness of the surfaces, which averages the surface forces, and also due to the low resolution of the FECO in CC (SM, **Figure S3**).

    The repulsion in CC is primarily a mechanical effect, reflecting the energy needed to compress multiple asperities upon loading. After the major recrystallization events, the range and onset of the repulsion increased, mainly because the topography of calcite surfaces was altered in a way that more asperities could come into contact. This was related to an overall increase in roughness (higher rms values) but also to the changes in asperity distribution: it was sometimes observed that the least numerous, highest asperities dissolved, and the distribution of surface heights became more symmetrical (see the histograms of heights of the calcite surface before and after the SFA experiment, **Figure S20**). Although we assume that these asperities were mainly deformed elastically, the pressures transmitted at the contacts could potentially lead to irreversible plastic deformation as well. Additionally, despite a normal load applied by the opposing surface, it can be imagined that there was some degree of friction between the asperities that mutually aligned in a contact in CC. Dominance of the mechanical effects was previously

found to strongly decrease or totally overcome the expected adhesive forces in numerous systems [43-44, 118-119, 131-135].

Repulsion in CC is also most likely influenced by chemical effects, including DLVO and non-DLVO forces. We evidenced no pronounced EDL repulsion nor VdW attraction. The former is weak but long ranged in the CaCO$_3$-saturated solutions ($\lambda_{Debye}$ ~ 16 nm) and its absence was related to roughness, which averaged a distribution of surface species, making the Stern layers more diffuse. The absence of VdW attraction might have been related to hydration force and/or roughness (due to elastic compression and insufficient contact areas). The repulsive forces between two smooth calcite surfaces [35] or between smooth calcite and silica surfaces [36] were only recently measured with AFM, and attributed to the repulsive hydration effects. Such hydration forces produce repulsive steric effects as they are related to a thin water layer, strongly adsorbed on a hydrophilic calcite surface [16]. They depend on the concentration of counterions, being stronger with more cations, but of a higher onset at their lower concentrations [36]. Therefore, we expect that hydration contributed to the repulsion measured in CC. It is possible that the absence of adhesion in case of the smoothest calcite surfaces could be attributed to this strongly repulsive effect. A larger range but lower magnitude of the hydration force is expected for rough surfaces, as only the highest asperities in contact will reach separations small enough to experience this nm-ranged interaction.

## Force Measurements Between Two Calcite Surfaces in MEG

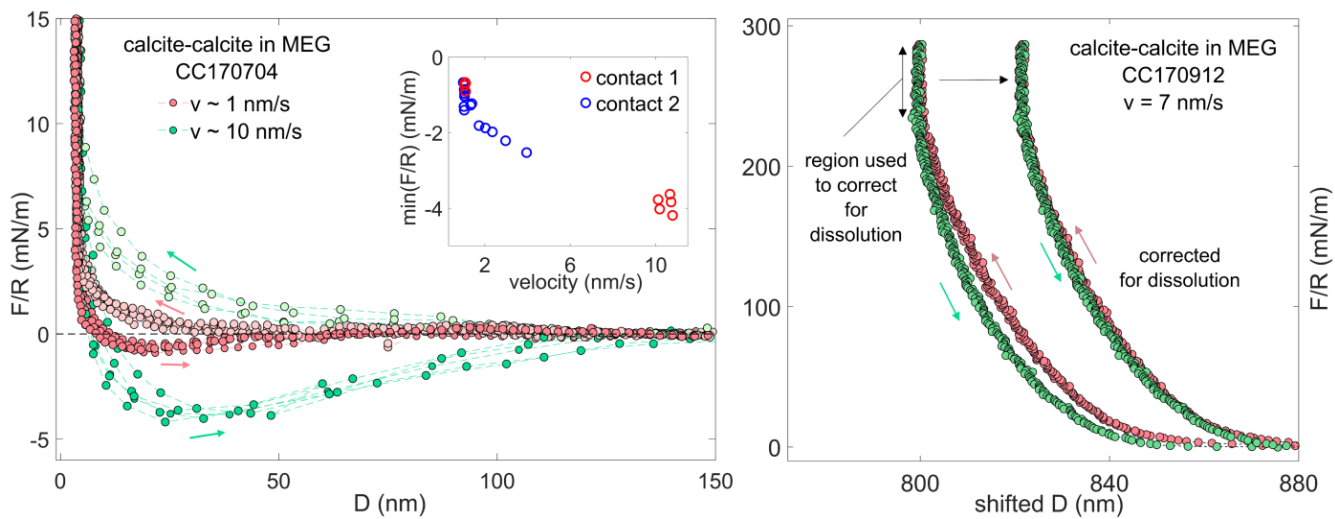

**Figure 13.** *Representative SFA force-distance measurements between: A) two relatively smooth calcite surfaces in MEG, showing velocity-dependent forces: a repulsive force on approach and an attractive force on retraction (in contact 1). The inset shows the minimum attractive force in each FR as a function of surface displacement velocity, measured in 2 different contact areas, using the same set of calcite surfaces (CC170704). The rms roughness measured after the experiment was 5.4 nm (scan size 20x20 µm$^2$); B) two rough calcite surfaces in MEG, showing a dissolution hysteresis between the FR measured on approach and retraction, and the same FRs corrected for dissolution, using the estimated dissolution rate determined from the marked region of the FR measured on retraction. The D values were arbitrarily shifted. The rms roughness measured after the experiment was 59 nm (scan size 50x50 µm$^2$).*

Forces between two calcite surfaces were additionally measured in MEG. The interactions in MEG were repulsive or attractive depending on the roughness of the surfaces: adhesion forces were present only for the relatively smooth calcite surfaces (CM170704), and exhibited a distinct dependence on velocity, being stronger with faster displacements **(Figure 13A)**. Both the magnitude and range of the repulsion measured on approach and the adhesion force measured on retraction increased with the velocity, giving rise to a velocity dependent hysteresis. That suggests a major influence of hydrodynamic forces. The minimum of the adhesion force measured on retraction in two different contacts was linearly proportional to the movement velocity of the surfaces **(Figure 13A, inset)**, as follows from the expression for the hydrodynamic force ($F_h$) between two crossed cylinders of radius R, at no-slip conditions: $F_h =$

$-\frac{6\pi\eta R^2 v}{D}$, where η is fluid viscosity, v is surface displacement velocity, and D is separation between the surfaces [136]. The action of roughness is to reduce hydrodynamic effects by increasing the degree of slip [137-138], explaining why no hydrodynamic force was measured for rougher CC surfaces **(Figure 13B).** Forces between rough CC surfaces in MEG were repulsive, and a hysteresis in the repulsive region of the force-distance curves was observed in all cases, analogously to the measurements in the $CaCO_3$-saturated solution. In case of the roughest surfaces, with <1 µm asperities (CC170912), limited but progressive dissolution was observed in MEG, uncorrelated to small changes in temperature **(Figure 13B;** SM, **Figure S19**). The observed hysteresis could be, in this case, almost entirely attributed to dissolution (using an estimated dissolution rate determined from the region of the smallest D values of the FR measured on retraction, as marked in **Figure 13B)**. Although adhesive forces has been previously measured between two calcite surfaces in MEG, and attributed to VdW attraction [35], it was not clear if the attraction measured in our system was also partially related to this interaction.

## Loading-Unloading Hystereses

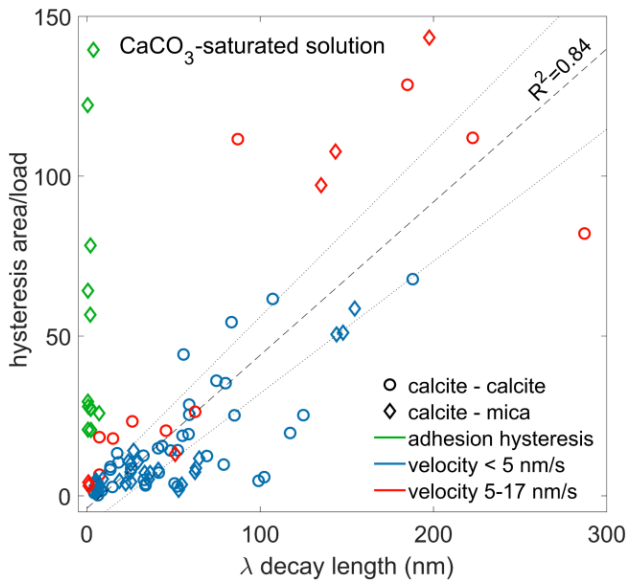

**Figure 14.** *Area of hysteresis between the FRs on approach and on retraction as a function of an exponential decay length λ (~ roughness) of the FRs on approach, for calcite-calcite (9 sets) and calcite-mica (7 sets) of surfaces. The areas of hystereses were normalized by the maximum applied load (mN/m) at which the hystereses were still present and were not corrected for calcite dissolution/growth. The green color corresponds to adhesive FRs between mica and calcite, the areas of which were calculated both in the attractive and repulsive region of the force-displacement curves. The red and blue color correspond to the repulsive FRs, with hystereses fully located in the repulsive region of the FR curves (as in **Figure 11**). The dashed line shows a linear regression fit without the attractive FRs. All the FRs were measured in $CaCO_3$-saturated solutions within 25 h from its injection.*

Areas of hystereses between the force-distance curves on approach and on separation can be, apart from the exponential decay length, another feature of the FR that provides information about a changing topography of the samples. We observed two types of hystereses in our system: adhesive hystereses for the attractive FRs in CM; and hystereses fully located in the repulsive region of the force-distance curves for the repulsive FRs in CC and CM. We argue that the repulsive hystereses observed in CC were mainly related to both mechanical effects during the loading-unloading cycles and to the changes in calcite thickness on dissolution and growth. The hystereses that would be produced by a mechanical backlash or thermal drifts would be significantly smaller than that the ones observed in our system.

The areas of hystereses between the FR on approach and on separation, normalized by the maximum applied load, were plotted in **Figure 14** as a function of the exponential decay length λ of the FR on approach (~estimated roughness in the contact area). We found that these areas were proportional to the roughness of the contacts established in the SFA: the larger the roughness magnitude, the larger the hystereses both in CC and CM. Additionally, areas of adhesion hystereses for the attractive FRs in CM were plotted, showing that the adhesion forces were measured only if λ was smaller than 8.5 nm. The hysteresis areas were affected by growth and dissolution of calcite because of the changes in calcite thickness and hardwall position: the areas were larger for dissolving surfaces and smaller growing surfaces, relative to no change in calcite thickness. In some cases, using an estimated constant dissolution rate, the hystereses could be fully attributed to calcite dissolution (**Figure 13B**; SM, **Figure S19**). In many cases, however, the large areas of hystereses could not be accounted only to dissolution or growth (**Figure 11**; SM, **Figures S20, S21**). We also measured FRs with no hystereses for relatively smooth calcite surfaces in CC (4<λ<9 nm), reflecting a dominance of elastic compression in these cases (e.g. SM, **Figure S21**, runs 4–10).

Hystereses in the force-distance curves in the SFA can arise due to adhesion forces, irreversible surface deformation (dissolution or plastic deformation), but also due to surface roughness. The adhesion hysteresis is related to a difference in the adhesion energy, which is larger on unloading (a contracting contact area) than on loading (a growing contact area)[16]. Hystereses related to the fast dissolution/growth occur because the sample thickness is changed during the loading-unloading cycles. Plastic deformation produces irreversible hystereses also because of the changes in a sample thickness but mostly because the energy needed to deform the surface plastically is not released back on unloading as in the case of a perfectly elastic smooth contact. Finally, hystereses in the repulsive region of the force-distance curves can be present for elastic contacts due to roughness of the surfaces [117, 139]. In such cases the number of asperities is higher on separation, and while some asperities still experience compression during the unloading, other shorter ones are stretched out and exert a tensile load on the opposing surface, decreasing the net load at a given separation. This behavior, however, has been evidenced in the case of attractive interaction between the rough surfaces [117, 139]. In our system, such mechanisms could possibly partially explain the repulsive hystereses in CM, when the pull-off forces were not measured due to high surface roughness. It is less likely that rough repulsive contacts (as in CC) between elastic surfaces would experience such interaction, as the small asperity junctions in this case can be described as a set of Hertzian-like repulsive contacts without the tensile stretching as predicted for the JKR-type adhesive contacts. Although we assumed before that the loading of the asperities was dominantly an elastic process, such process should not produce the sometimes observed large hystereses, especially at times when dissolution of surfaces was limited. The previously calculated PI values suggested that plastic deformation was likely for our rough calcite films. The PI parameter implies that plastic deformation is easier for smaller asperities, and that deformation degree is higher when there are more asperities on the surface. This agrees with the trend observed in **Figure 14**, showing larger hystereses for the rougher surfaces having more asperities. In the experiments in which calcite films underwent significant recrystallization and roughening overnight (CM170711, **Figure 6**; CC170713, **Figure 7**), the observed hystereses seem to be clearly indicative of subsequent plastic deformation of the roughened films (SM, **Figure S20**, runs 9–14; **Figure S21**, runs 12–18): the areas of hystereses decreased in the consecutive FRs and the HP shifted to lower values, but the FR on separation did not show any progressing dissolution of the surfaces (as in **Figure 13B** where the FR on retraction shifts to smaller D values despite moving the surfaces out). When the surfaces were more reactive, resulting in faster HP shifts, it was difficult to separate dissolution and the HP shifts due to plastic deformation, as the dissolution rates were not constant and could be very easily misestimated. Similarly, we could not observe a major reduction of hystereses in the repeated loading-unloading cycles for more reactive surfaces (which should happen as the degree of irreversible deformation would be gradually smaller). Highly likely, the plastic deformation still occurred at the asperity tips in most cases, but the contact topography was affected by recrystallization in the same time, 'renewing' the contacts for plastic deformation. Gradually reduced but not completely eliminated hystereses have been previously observed on repeated loading-unloading for plastically deforming asperities, which exhibited some recovery of deformation when the surfaces were separated [140]. We thus suggest that in our system, in the case of repulsive contacts with no pronounced growth or dissolution, λ of the FR during loading could be interpreted as due to both elastic and plastic deformation of the asperities in the contact, and λ of the FR during unloading would be indicative of the magnitude of elastic compression.

It is also possible that when two rough surfaces were contacted, the asperities aligning in contact and separating from the established contact experienced friction. This could be reflected in FRs on approach where more energy would be needed to bring such rough surfaces into contacts, and in FRs on separation where friction would oppose the surface retraction, adding to the hysteresis. However, such contribution would not be present in CM, in which system one surface was atomically smooth.

## Roughness Modelling

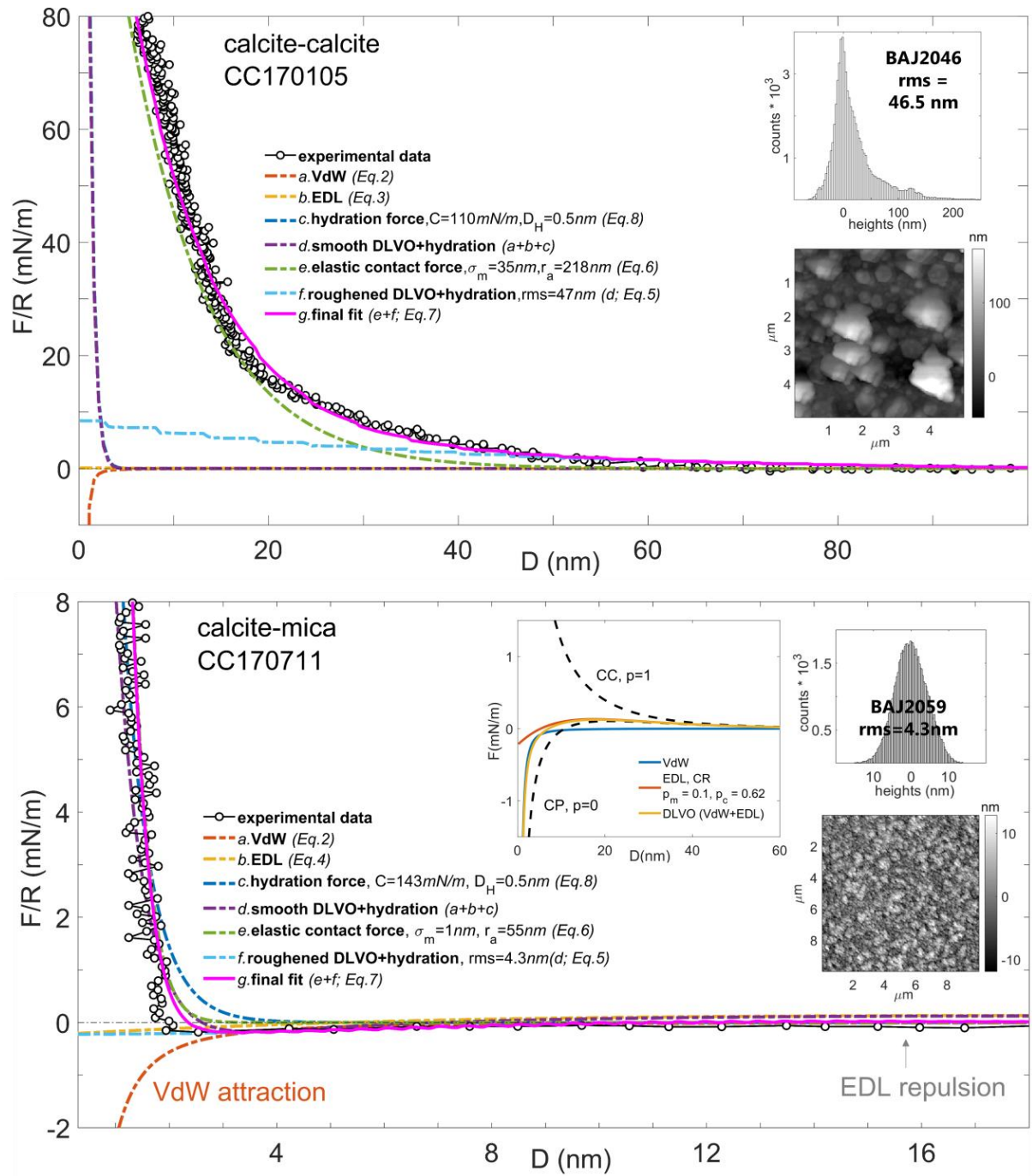

**Figure 15.** *Modelling of the roughness contribution to the total forces between two calcite (top) and calcite and mica (bottom) surfaces. The final fit (solid line) was a sum of the elastic contact force and the roughened DLVO + hydration forces (Eq.7). The insets show histograms of heights and topographies of the corresponding calcite surfaces measured by AFM and a zoom on the theoretical DLVO forces between smooth calcite and mica surfaces (Eq.2,4). In CC, c. and d. contributions are overlapped.*

It is frequently difficult to distinguish between the repulsive force contributions of surface roughness and hydration forces, because both add an exponentially repulsive component to the total force acting between the surfaces. In order to estimate the roughness effect on the total forces in CC and CM, we used a model incorporating two components: 'roughened' DLVO forces and a repulsive force due to compression of surface asperities [44-45]. In the first component, roughness-averaged DLVO forces (Eq. 5), separation values between the surfaces are weighed against the probability distribution of the surface heights: the DLVO potential is 'smeared out', resulting in an onset at larger average separation, but a smaller magnitude of the force at a given separation, than for atomically smooth surfaces. The interpretation of the roughness-averaged forces is strongly affected by the definition of the hardwall (0 nm separation) between the rough surfaces [141]. For our data, the hardwall position was shifted so that the separation is 0 nm at the maximum applied load in each FR. Due to the uncertainty in absolute separation between the surfaces, the onset distances of DLVO and hydration forces were likely to be misestimated. The second model component (Eq. 6) accounts for the mechanical work that is needed to compress the surface asperities elastically, and thus always produces a repulsive force. The total interaction is defined as the sum of these two contributions (Eq. 7), with iteratively adjusted surface separations [45].

The roughness contribution to the total forces was first estimated in CC in a $CaCO_3$-saturated solution **(Figure 15).** The rms roughness of the calcite films in the exemplary FR (CM170195) was 46 nm (scan size 5x5 $\mu m^2$) and the mean radius of the highest asperities (30%) was 218 nm, as measured with AFM (these values are approximations based on force measurements in air, using a different position on the surfaces than subsequent measurements in liquid, but we expect rms in the contact region to be in the same range). The theoretical DLVO forces in this system predicted a very small-magnitude EDL repulsion (Eq. 3), and a strong, short-range VdW attraction (Eq. 2). Instead, we measured a long-range repulsion with an onset at approximately 70 nm from the estimated HP, and the magnitude 100 times higher than the theoretical EDL forces. The roughness-averaged DLVO forces (Eq.5) showed a longer range and a lower magnitude attractive force. The contact force (Eq. 6) was strongly repulsive due to high calcite roughness. Assuming the rms roughness of both calcite surfaces to be 46 nm (as measured with the AFM), the high-magnitude experimental repulsion could be solely explained by the repulsive contact force component (Eq. 6). Even though the experimental forces could be well modelled by the contact repulsion only, it is highly likely that additional repulsive hydration forces acted between the surfaces [35-36]. All calcite surfaces are hydrated. When such surfaces are brought into contact, repulsive effects due to a progressive dehydration of surfaces of the highest asperities upon an increasing load, can further enhance a magnitude of the repulsive contact force. In theory, the hydration forces are reasonably described by an exponential term with a small decay lengths of a few nm, which has been measured for silica surfaces and smooth mica surfaces in different electrolyte solutions [17, 142-143]. Such exponentially repulsive hydration forces have been characterized by decay lengths varying between 0.5 nm for silica and maximum 2 nm for mica. As the used model originally included only the DLVO interaction, we added a simplified hydration potential to the DLVO interaction energy for the smooth surfaces, and averaged the sum of these contributions for roughness (Eq. 5). To account for the hydration, we used a force law in a form:

$$\frac{F}{R} = C \cdot \exp(-\frac{D}{D_H}) \qquad (Eq. 8)$$

where $C$ is an empirical force constant (N/m), and $D_H$ is the hydration decay length (m) [17, 143]. The final fit (Eq. 7) was given as a sum of such 'roughened' DLVO + hydration and the repulsive compression force calculated for the smaller rms roughness of 35 nm than the rms measured with the AFM (46 nm). We assumed a decay length $D_H$ of the smooth hydration force of 0.5 nm and the force constant $C$ of 110 mN/m. When the hydration repulsion was included, the contribution of the contact force due to deformation could be accordingly smaller. Because of the uncertainty in rms values at the contact established in the SFA, it was not possible to precisely distinguish the contributions of the hydration forces and the contact forces, since both interactions are exponentially repulsive. Nevertheless, according to the modelling, most of the repulsive effects in this system could be attributed to a mechanical work needed to compress the surface asperities, and this repulsion was likely to be enhanced by the hydration force to some extent.

Roughness contribution was also modelled in CM in $CaCO_3$-saturated solution **(Figure 15)**, using an exemplary FR (CM170711) showing a small attractive force on approach. The rms roughness of the single calcite surface was much smaller – 4.3 nm (scan size 10x10 µm$^2$), and the mean radius of the highest asperities (15 %) was approximately 55 nm, as measured with the AFM. In order to model an atomically flat mica surface with no asperities, we set the rms value of to be 0.05 nm and the radius of asperities to be very large (150 µm). No major repulsion, either due to elastic deformation or EDL, was experimentally measured, although, in theory, a small-magnitude repulsive EDL force (Eq. 4) should be present in our system **(Figure 15, inset).** The EDL absence was related to the roughness, which caused the distribution of surface species to be more diffuse, as suggested by the roughness-averaged DLVO forces (Eq. 5), showing no well-defined EDL region. Based on the AFM-measured rms, we predicted a repulsive contact force (Eq. 6) which was not measured experimentally. By assuming a lower rms of 1 nm, we obtained a more correct repulsion magnitude in this case. Using this roughness value, the 'roughened' DLVO component and the contact force could generate a good fit to the experimental data with the iteratively adjusted onsets for each contribution (not shown). However, it is expected that hydration force acted between the CM surfaces as well. The measurements between two mica surfaces in $CaCO_3$-saturated solutions did not show any high-magnitude hydration effects (SM, **Figure S14**), which could be explained by the fact that the hydration of mica is a secondary-type (related to dehydration of surface-adsorbed cations), and in low-concentration electrolytes (0.12–0.17 mM) used in our experiments, the hydration effects related to mica were weak [17,142]. Thus, the hydration forces should have been mostly related to the calcite surface. The final fit including the hydration force calculated in the same way as in CC is shown in **Figure 15**. The used values of $C$ and $D_H$ were only exemplary and depended mostly on the onset chosen for the 'roughened' DLVO interaction. It was possible that other non-DLVO effects, such as ion correlation forces, acted also between the surfaces. Nevertheless, as follows from the modelling for the smooth CM surfaces, the contact force component was negligible here, in contrast to the example with much rougher calcite-calcite surfaces.

## Implications

The crucial role of interfacial fluids in deformation of porous carbonate sediments is manifested in much higher rates of chemical compaction than of mechanical compaction. The high reactivity and solubility of $CaCO_3$, relatively to silicate and aluminosilicate minerals, and high porosity of carbonate rocks makes them especially prone to phenomena such as the water-weakening [20]. In our system, we generally observed two processes with an opposite effect on the strengthening of solid-solid interfaces: 1) nm-scale recrystallization in water, leading to relatively strong repulsive effects between the surfaces; 2) increase in contact area, and thus the adhesion force between the surfaces.

Firstly, the experimentally observed lower mechanical strength of water-saturated carbonate rocks, such as chalk, at very short time scales, has been attributed to the loss of cohesion between the individual calcite grains in water [15]. This relatively fast loss of cohesion has been linked with various mechanical and chemical mechanisms [13, 15, 22, 24, 35, 144-146]. However, the possible increase of calcite roughness upon equilibration with pore fluids of varied chemistry, even at a nm-scale, has rarely been discussed in relation to the water-weakening. We showed that the confined calcite surfaces can recrystallize and grow even when the bulk solution is undersaturated with respect to calcite. The mechanical effects related to nm-scale recrystallization of calcite surfaces could cause the observed repulsive force to be of much higher magnitude and onset than the EDL repulsive forces and even the strongly repulsive hydration forces. We thus propose that the nm-scale recrystallization of calcite surfaces at solid-solid contacts in water can additionally cause a significant decrease in cohesion between calcite grains. Assuming that the real contact area in our system was only a small fraction of the nominal contact area, the force of crystallization associated with the nm-scale recrystallization of calcite crystals could overcome the confining pressures of the order of MPa. In addition, the previously observed chalk strengthening in MEG by Risnes, et al. [15], could be also partially related to the less pronounced roughening of calcite in MEG observed in our system.

Secondly, we showed that the gradual plastic deformation of calcite asperities in contact with mica was most likely responsible for the observed smoothening of contact areas and the increase in pull-off forces, and thus the strengthening of the interface between these two minerals. Such phenomenon can have a major significance for an increased resistance of frictional interfaces [147].

Moreover, the observed increase of calcite roughness in water can be significant for EOR in chalk and many industrial systems, as apart from enhancing the repulsive forces between contacting calcite grains, it also affects the surface wettability properties [42].

# Conclusions

We showed that the topographical evolution of rough calcite surfaces depended on the degree of confinement, with rougher contacts being less isolated from the bulk fluid, and undergoing more pronounced dissolution. Prolonged dissolution and limited mass transport between the smoothest sets of surfaces could lead to increasing supersaturation in the gap between the surfaces, calcite growth, major roughening, and the force of crystallization that could overcome confining pressures of the order of MPa. The changes in surface roughness were closely related to the forces acting between the surfaces. The rougher the surfaces, the higher the magnitude of the repulsive contact force, which was related to mechanical work needed to compress multiple asperities on loading. The repeated loading-unloading cycles in the case of adhesive systems led to a gradual increase in pull-off forces due to increasing contact areas, most likely caused by a progressive plastic deformation of asperities on the rough calcite surfaces. Only repulsive forces between even the smoothest two calcite surfaces were measured, possibly related to repulsive hydration effects. The relatively strong adhesive interaction between smoother calcite and atomically smooth mica surfaces was possibly enhanced by electrostatic effects. We propose that surface roughening of calcite in water could be another mechanism that enhances the water-weakening effect in carbonates by decreasing cohesion between calcite grains and facilitating grain sliding upon compaction.

# ACKNOWLEDGEMENTS

This project has received funding from the European Union Horizon 2020 research and innovation program under the Marie Skłodowska-Curie grant agreement no. 642976-NanoHeal Project. This work reflects only the author's view and the Commission is not responsible for any use that may be made of the information it contains. SJ acknowledges the National IOR center of Norway under project no. PR-10373. The AFM measurements were perfomed by the JKR system funded by B. Jamtveit from the European Union's Horizon 2020 research and innovation programme under the ERC advanced Grant Agreement (669972). B. Løken Berg is acknowledged for the help with SEM. G. Jonski is acknowledged for the assistance with AAS. We thank L. de Ruiter and O. B. Karlsen for the useful advice.